\documentclass[pdflatex,sn-mathphys-num]{sn-jnl}

\usepackage{array}
\usepackage{pdflscape}
\usepackage{longtable}
\usepackage{graphicx}%
\usepackage{multirow}%
\usepackage{amsmath,amssymb,amsfonts}%
\usepackage{amsthm}%
\usepackage{mathrsfs}%
\usepackage[title]{appendix}%
\usepackage{xcolor}%
\usepackage{textcomp}%
\usepackage{manyfoot}%
\usepackage{booktabs}%
\usepackage{algorithm}%
\usepackage{algorithmicx}%
\usepackage{algpseudocode}%
\usepackage{listings}%
\usepackage{float}
\usepackage{bm}

\providecommand{\AUC}{\operatorname{AUC}}
\providecommand{\KL}{\operatorname{KL}}
\providecommand{\TV}{\operatorname{TV}}
\providecommand{\E}{\mathbb{E}}
\providecommand{\Prob}{\mathbb{P}}
\theoremstyle{thmstyleone}%

\theoremstyle{thmstyletwo}%

\theoremstyle{thmstylethree}%

\begin{document}

\title[]{Energy, Information, and the Limits of Neuropsychiatric Classification}

\author[1]{\fnm{Farbod} \sur{Khodadadi}}
\author[2,3]{\fnm{Mina} \sur{Kheirkhah}}
\author[1]{\fnm{Bita} \sur{Shariatpanahi}}
\author[1,5]{\fnm{Sarah} \sur{Alizadeh}}
\author[1]{\fnm{Madleen} \sur{Stenger}}
\author[1]{\fnm{Amaia} \sur{Zurinaga Gutierrez}}
\author[1]{\fnm{Shakiba} \sur{Bakhshian}}
\author[1]{\fnm{Jos\'e C.} \sur{Garc\'ia Alanis}}
\author[4]{\fnm{Nils R.} \sur{Winter}}
\author[1]{\fnm{FOR2107} \sur{Consortium}}
\author[5,6,7]{\fnm{Frank} \sur{Bremmer}}
\author[2]{\fnm{Stefan G.} \sur{Hofmann}}
\author[4,8,9,10]{\fnm{Udo} \sur{Dannlowski}}
\author[1]{\fnm{Tilo} \sur{Kircher}}
\author[11]{\fnm{Sharmili} \sur{Edwin Thanarajah}}
\author[1]{\fnm{Igor} \sur{Nenadi\'c}}
\author[4]{\fnm{Tim} \sur{Hahn}}
\author*[1,5]{\fnm{Hamidreza} \sur{Jamalabadi}}\email{hamidreza.jamalabadi@uni-marburg.de}

\affil[]{\normalfont\fontsize{9}{11}\selectfont\textbf{FOR2107 Consortium:} Frederike Stein\textsuperscript{1}, Florian Thomas-Odenthal\textsuperscript{1}, Paula Usemann\textsuperscript{1}, Lea Teutenberg\textsuperscript{1}, Benjamin Straube\textsuperscript{1}, Nina Alexander\textsuperscript{1}, Andreas Jansen\textsuperscript{1}, Nooshin Javaheripour\textsuperscript{1}, Susanne Meinert\textsuperscript{4}, Elisabeth J. Leehr\textsuperscript{4}, Tiana Borgers\textsuperscript{4}, Kira Flinkenfl\"ugel\textsuperscript{4}, Dominik Grotegerd\textsuperscript{4}, Lukas Fisch\textsuperscript{4}, Julia Hubbert\textsuperscript{4}, Christoph Jurischka\textsuperscript{4}, Judith Krieger\textsuperscript{4}, Anne Hildebrand\textsuperscript{4}, Roman Vogler\textsuperscript{4}, Sabrina Teckentrup\textsuperscript{4}, Sara Papenbrock\textsuperscript{4}, Franziska Herrmann\textsuperscript{4}, Elilarasi Ratnalingam\textsuperscript{4}, Theresa Slump\textsuperscript{4}, Hannah Meinert\textsuperscript{4}.}

\affil[1]{\orgdiv{Department of Psychiatry and Psychotherapy}, \orgname{Marburg University}, \country{Germany}}
\affil[2]{\orgdiv{Department of Psychology}, \orgname{Marburg University}, \country{Germany}}
\affil[3]{\orgdiv{Department of Psychiatry and Psychotherapy}, \orgname{Jena University Hospital}, \country{Germany}}
\affil[4]{\orgdiv{Institute for Translational Psychiatry}, \orgname{University of M\"unster}, \country{Germany}}
\affil[5]{\orgdiv{Center for Mind, Brain, and Behaviour (CMBB)}, \orgname{Marburg University}, \country{Germany}}
\affil[6]{\orgdiv{Department of Physics}, \orgname{Marburg University}, \country{Germany}}
\affil[7]{\orgdiv{Department of Physiology}, \orgname{Monash University, Melbourne}, \country{Australia}}
\affil[8]{\orgdiv{Department of Psychiatry}, \orgname{Medical School and University Medical Center OWL, Protestant Hospital of the Bethel Foundation, Bielefeld University}, \country{Germany}}
\affil[9]{\orgname{German Center for Mental Health (DZPG), Site Jena Magdeburg Halle}, \country{Germany}}
\affil[10]{\orgname{Center for Intervention and Research on Adaptive and Maladaptive Brain Circuits Underlying Mental Health (C-I-R-C), Site Jena Magdeburg Halle}, \country{Germany}}
\affil[11]{\orgdiv{Department of Psychiatry, Psychosomatics and Psychotherapy}, \orgname{Goethe University, Frankfurt am Main}, \country{Germany}}

\abstract{%
\noindent\textbf{Importance}\par
\noindent The search for reliable neuroimaging biomarkers of mental disorders has remained elusive despite decades of research. Why clinically consequential disorders yield weak neuroimaging separability remains unresolved. Drawing on statistical thermodynamics and the metabolic costs of neural coding, we propose that psychiatric disorders produce constrained deformations of a largely shared repertoire of brain activity patterns, limiting the class-specific information available for classification.
\par\vspace{0.6em}
\noindent\textbf{Observations}\par
\noindent We develop a framework linking physical resource constraints, landscape deformation, distributional separation, and attainable classification, and examine its components using human neuroimaging, synthetic data, and an exactly specified physical model. In resting-state fMRI from 1952 FOR2107 participants (823 with major depressive disorder; 1129 healthy controls), held-out energy-landscape estimates and nonlinear classifiers agreed across 13 comparisons ($r = 0.996$). Similar results emerged at both FOR2107 centers and in 900 SRPBS participants spanning multiple psychiatric diagnoses. In synthetic data, classification improved with distributional separation and approached the optimum. In an equilibrium three-well model, statistical energy equaled physical potential energy scaled by inverse temperature. Modest deformation redistributed state occupancy while optimal snapshot classification remained weak. A literature map found insufficiently comparable evidence to test the human metabolic constraint.
\par\vspace{0.6em}
\noindent\textbf{Conclusions}\par
\noindent Clinically consequential differences may arise within a largely shared neural repertoire. The thermodynamic framework connects physical deformation to classification under specified assumptions; its metabolic implementation in humans remains untested. Testing this mechanism requires joint measurements of brain activity, bodily physiology, and metabolism.}


\maketitle

\section{Introduction}
\label{sec:introduction}

Psychiatric disorders can profoundly alter mood, cognition, motivation, and behavior \citep{marx2023major,american2022diagnostic}. Yet reliable neuroimaging biomarkers remain elusive despite decades of intensive research. Even large, harmonized studies lack individual-level markers with clinically relevant accuracy. Structural and functional correlates are typically weak, heterogeneous, and extensively overlap between affected and unaffected individuals \citep{winter2022quantifying}. In major depressive disorder (MDD), analysis of 4,021 individuals from three clinical cohorts found weak but replicable neuroanatomical effects \citep{dannlowski2026neuroanatomy}. Large multisite studies likewise report modest average neuroimaging effects, and diagnostic performance remains similarly modest despite varied preprocessing, features, modality, and architecture \citep{winter2024systematic}. Independent multicenter functional and structural imaging analyses concur \citep{gallo2023functional,belov2024multi}.

This plateau---the failure to recover a reliable individual diagnosis---raises a question more fundamental than classifier choice: how much diagnostic separation exists in the measured distributions? The mathematical limit is simple. For a fixed measurement and representation---for example, adults aged 20--65 years with depression, imaged with BOLD fMRI and summarized by functional connectivity---the class-conditional distributions determine the best attainable discrimination. Ranking observations by the likelihood ratio yields the optimal ROC curve \citep{neyman1933ix}. A flexible decoder may approach that optimum, but cannot recover class information absent from its input. Achieved performance therefore depends on sample, estimation, and algorithm; the population-level limit depends on the representation---and its underlying biology \citep{woo2017building}.

That limit does not explain large clinical differences despite overlapping images. Dynamical-systems accounts offer one biological explanation \citep{scheffer2024dynamical,kheirkhah2026reengineering}. Psychiatric disorders need not create distinct neural architectures or entirely new brain-state repertoires. Small geometric or flow changes in a shared dynamical system can alter state occupancy, persistence, transition probabilities, and perturbation responses \citep{deco2011emerging,breakspear2017dynamic}. Because behavior depends on these temporal properties, these changes can profoundly affect function while only modestly altering static or cross-sectional representations. Consistently, MDD studies report altered temporal variability, occupancy, and transitions of functional brain states \citep{long2020altered}.

\begin{figure}[H]
    \centering
    \includegraphics[width=\textwidth]
    {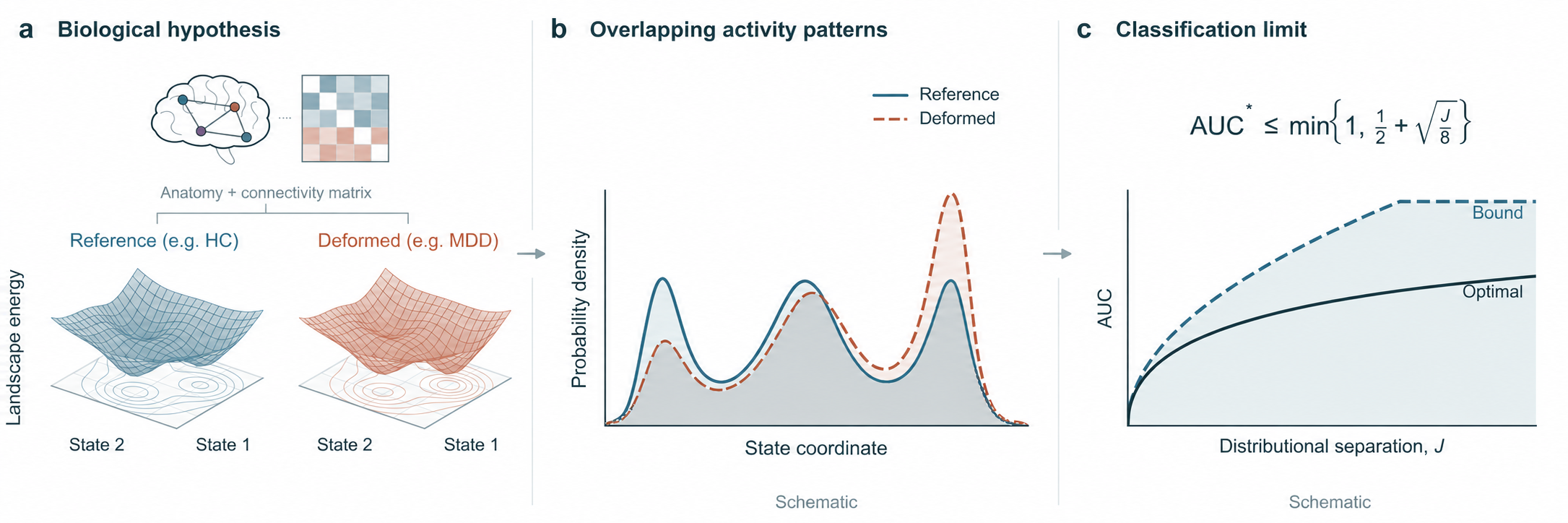}
    \caption{\textbf{Proposed framework linking physical
    constraints to neuroimaging classification limits.}
    \textbf{A,} Connectomic architecture and finite physical
    resources are hypothesized to constrain how inputs and
    experience deform a shared neural activity landscape.
    \textbf{B,} Deformations can redistribute state occupancy
    while leaving measured class-conditional distributions
    substantially overlapping.
    \textbf{C,} Distributional separation limits optimal
    snapshot classification according to
    $\mathrm{AUC}^{*}\leq
    \min\{1,\frac{1}{2}+\sqrt{J/8}\}$.
    Neuroimaging tests the distributional consequence;
    synthetic experiments evaluate classification against
    known source distributions; and the mechanical system
    connects physical deformation, protocol-specific work,
    occupancy, separation, and optimal classification under
    equilibrium assumptions. Statistical energy in the
    neuroimaging model is probability-based and does not
    measure metabolic expenditure. The proposed human
    resource constraint remains a biological hypothesis.}
    \label{fig:theoretical_framework}
\end{figure}

Drawing on statistical thermodynamics, metabolic costs of neural coding, and the classification bound above, we hypothesize that psychiatric disorders---examined primarily through MDD in FOR2107 and secondarily through multi-disorder comparisons in SRPBS---produce energy-limited deformations of a largely shared activity landscape constrained by connectomic architecture (Section~\ref{sec:theoretical_framework}; Fig.~\ref{fig:theoretical_framework}). Neural representations are physical: membrane potentials, spikes, synaptic activity, and circuit modification require energy \citep{attwell2001energy,jamadar2025metabolic}. The brain encodes sensory, interoceptive, and social regularities with finite metabolic supply. In this proposal, input reshapes a shared repertoire, connectomic organization restricts activity propagation directions, and metabolism limits the extent, speed, and stability of those deformations \citep{jaynes1957information,ezaki2017energy}. The constraint is not globally scalar. Energy demand differs across neural populations, and circuits for cognitive control and effort-based decision-making are particularly costly, so reduced energy availability should tilt the landscape unevenly \citep{jamadar2025metabolic}. Peripheral and interoceptive signals further modulate node-specific energy status, coupling the accessible repertoire to organismal homeostasis rather than to a purely central budget \citep{mehrhof2025interoceptive}. Behavior emerges from supported trajectories; neuroimaging incompletely observes their distribution. Both datasets provide one time-averaged connectivity snapshot per person; fitted distributions therefore describe group-level snapshot overlap.

We test this distributional consequence in four steps. First, in functional-connectivity data from 1,952 participants in the Marburg--M\"unster FOR2107 cohort \citep{kircher2019neurobiology}, we compare separability hierarchies from held-out maximum-entropy likelihood-ratio estimates and nonlinear classifiers across thirteen age, recorded-sex, MDD, and symptom contrasts, and test replication in each independently acquired center. We repeat this comparison in 900 participants from the multisite SRPBS dataset, which includes MDD, autism spectrum disorder, schizophrenia-spectrum disorder, and healthy controls \citep{tanaka2021multi}. Second, in controlled simulations with known source distributions, we test whether landscape separation determines attainable AUC and whether nonlinear classifiers approach the source-defined optimum. Third, an exactly solvable shared triple-well model examines how modest deformations can redistribute state occupancy while snapshot classification remains weak. Fourth, we review the metabolic and neuroimaging-classification literatures \citep{kheirkhah2026targeted} to assess whether human evidence can test the link between physical energy budgets and classification limits. These analyses distinguish established distributional findings from the biological mechanism requiring multimodal experiments.

\section{Theoretical framework}
\label{sec:theoretical_framework}

We propose that psychiatric disorders alter psychological function through resource-constrained deformations of a largely shared neural repertoire. Dynamical accounts motivate changes in stability, occupancy, and transitions \citep{scheffer2024dynamical,kheirkhah2026reengineering}; neural models explain how connectivity, coupling, and noise shape this repertoire \citep{deco2011emerging,breakspear2017dynamic,suarez2020linking}. Altered functional-network dynamics and co-activation probabilities in depression motivate examining its statistical expression \citep{long2020altered,teutenberg2025synergistic}. The metabolic costs of signaling and cognition motivate, without establishing, a physical constraint on these deformations \citep{attwell2001energy,jamadar2025metabolic}:
\begin{equation}
\begin{aligned}
&\text{architecture and resource constraints}\\
&\quad\longrightarrow\text{accessible landscape deformations}\\
&\quad\longrightarrow\text{distributional separation}\\
&\quad\longrightarrow\text{attainable classification}.
\end{aligned}
\label{eq:framework_chain}
\end{equation}

Let $p_\theta(\bm z)$ describe a fixed measurement and representation under landscape parameters $\theta$. Maximum-entropy landscapes express these probabilities through statistical energies \citep{jaynes1957information,ezaki2017energy}. Separation is quantified by
\begin{equation}
J(p_0,p_1)=D_{\mathrm{KL}}(p_0\Vert p_1)
+D_{\mathrm{KL}}(p_1\Vert p_0).
\label{eq:framework_divergence}
\end{equation}
Locally, $J=\delta\theta^\top I(\theta)\delta\theta
+O(\|\delta\theta\|^3)$, where $I$ is Fisher information. Assuming a quadratic accessibility cost
$\delta\theta^\top C\delta\theta\leq B$, with $C\succ0$, gives
\begin{equation}
J\leq B\Lambda+O(B^{3/2}),\qquad
\Lambda=\lambda_{\max}
\!\left[C^{-1/2}I(\theta)C^{-1/2}\right].
\label{eq:framework_resource}
\end{equation}
$B$ is the budget and $\Lambda$ the greatest local distributional change per unit cost. This conditional bound accommodates unequal deformation costs; its metabolic interpretation requires independent physical calibration (Appendix~\ref{app:resource_bound}).

For $p_g(\bm z)=Z_g^{-1}e^{-E_g(\bm z)}$, likelihood-ratio ranking is equivalent to the energy-difference score \citep{neyman1933ix,devroye1996probabilistic}:
\begin{equation}
q(\bm z)=E_0(\bm z)-E_1(\bm z),\qquad
\AUC^*\leq
\min\left\{1,\frac12+\sqrt{\frac{J}{8}}\right\}.
\label{eq:framework_bounds}
\end{equation}
The inequality is generally loose; known distributions determine exact optimal AUC (Appendices~\ref{app:optimal_ranking} and \ref{app:auc_bound}).

Statistical energy is negative log probability up to an additive constant, not glucose use, ATP turnover, oxygen consumption, or heat. Under an equilibrium Boltzmann model it corresponds to $\beta U$, with physical energy $U$ and inverse temperature $\beta$. This identification cannot simply be transferred to the open, driven, nonequilibrium brain \citep{jamalabadi2026metabolic}. We therefore distinguish distributional tests from the stronger biological hypothesis that physical resources constrain accessible landscapes.

In FOR2107 \citep{kircher2019neurobiology}, we test whether held-out generative scores and nonlinear classifiers recover the same separability hierarchy across thirteen contrasts. We repeat the comparison across fourteen demographic, diagnostic, and symptom contrasts in the independent SRPBS multi-disorder dataset \citep{tanaka2021multi}. Each observation summarizes one participant's time-averaged connectivity. The 120-dimensional binary representation and tree-structured model \citep{chow1968approximating} estimate separability within that representation; they do not define a universal ceiling for neuroimaging classifiers. Synthetic experiments vary known source distributions to test whether classifiers approach their exact optimum.

Finally, a mechanical system specifies the physical-to-statistical chain:
\begin{equation}
U_\alpha(x)=ax^2(x^2-1)^2-\alpha hx,\qquad
p_\alpha(x)=Z_\alpha^{-1}e^{-\beta U_\alpha(x)}.
\label{eq:framework_mechanical}
\end{equation}
Its shared triple-well structure permits joint evaluation of deformation, occupancy, work, separation, and AUC. For reciprocal instantaneous switches, each starting at equilibrium at the same temperature \citep{kawai2007dissipation},
\begin{equation}
J(p_0,p_\alpha)=
\beta\left(
W_{\mathrm{diss}}^{0\rightarrow\alpha}
+W_{\mathrm{diss}}^{\alpha\rightarrow0}
\right).
\label{eq:framework_work}
\end{equation}
This evaluates the deformation--work--distribution--classification chain under specified physical assumptions (Appendices~\ref{app:reciprocal_work} and \ref{app:triple_well}), including whether occupancy redistribution coexists with weak classification; switching work is not neural maintenance expenditure. This framework is testable; validating its human metabolic mechanism requires joint neural, connectomic, and physiological measurements.

\section{Results}
\label{sec:results}

\subsection{Generative and discriminative analyses recover the same hierarchy of neuroimaging separability}
\label{sec:results_empirical}

The same 120-variable representation, tree-structured maximum-entropy model, and RBF support-vector classifier were applied to two resting-state cohorts. FOR2107 is the primary sample; the smaller, independent multi-disorder SRPBS sample tests generality of the hierarchy.

\subsubsection{Marburg--M\"unster FOR2107 cohort}
\label{sec:results_for2107}

We tested the framework's distributional consequence in 1,952 FOR2107 participants: 1,129 healthy controls and 823 participants with MDD. Each participant contributed 4,950 functional connections among 100 brain regions. Thirteen binary comparisons covered recorded sex, age thresholds of 30, 40, and 50 years, healthy-control versus MDD status, and eight diagnosis-stratified contrasts involving BDI-I depression severity or SCL-90-R global symptom burden. Within-diagnosis medians defined sample-relative symptom strata rather than clinical cutoffs.

Representation learning and model fitting used only training folds in five repeats of stratified five-fold cross-validation. Connectivity was compressed into a 120-dimensional binary representation, and class-specific maximum-entropy distributions were fitted on a common maximum-information spanning tree. Fitted statistical-energy differences ranked held-out participants. We compared this generative score with an RBF support-vector classifier trained on the same fold-specific representation (Appendices~\ref{app:empirical_methods} and \ref{app:maxent_methods}).

Both approaches recovered nearly identical classification-difficulty hierarchies (Fig.~\ref{fig:main_results}A). Across the thirteen contrasts, mean AUC was $0.661$ for maximum entropy and $0.669$ for the RBF classifier, with an outcome-level correlation of $r=0.996$. Age was most distinguishable: maximum-entropy AUCs were $0.770$, $0.804$, and $0.796$ for the three thresholds. Recorded-sex AUC was $0.714$, versus $0.613$ for healthy-control versus MDD classification and $0.598$ to $0.628$ for symptom-defined comparisons. Fold-level estimates retained this demographic-versus-clinical ordering.

Held-out marginal agreement supported the models' capture of first-order and pairwise structure in the binary representation (Appendices~\ref{app:marginal_diagnostics} and \ref{app:heldout_adequacy}; Appendix Fig.~\ref{fig:pooled_marginal_agreement}). Analyses repeated separately in the non-overlapping Marburg and M\"unster subsets also reproduced the broad hierarchy and close generative--discriminative agreement, despite independent acquisition by different teams and scanners (Appendix~\ref{app:site_replication}; Appendix Fig.~\ref{fig:site_replication}a,c).

These results support the distributional argument: clinical contrasts remained weakly distinguishable with either generative or discriminative estimation. Nevertheless, fitted maximum-entropy AUC was a conditional reference. The RBF classifier slightly exceeded it for some outcomes, and classifiers using uncompressed connectivity exceeded it further (Appendix~\ref{app:supplementary_results}). The finding therefore establishes agreement in separability hierarchies within the analyzed representation, not a universal classification ceiling.

\subsubsection{SRPBS multi-disorder dataset}
\label{sec:results_srpbs}

To test the generality of FOR2107 agreement, we repeated the 120-variable analysis in 900 participants from the SRPBS Multi-disorder MRI Dataset: 500 healthy controls, 220 participants with MDD, 125 with autism spectrum disorder (ASD), and 55 with schizophrenia-spectrum disorder (SSD). Each analysis stage used only cross-validation training data. The analysis included recorded sex, the same three age thresholds, all six pairwise diagnostic comparisons, and four BDI-II comparisons between healthy controls and participants with MDD, divided into high- and low-symptom groups (Appendix~\ref{app:srpbs_results}).

\begin{figure}[H]
    \centering
    \includegraphics[width=\textwidth]
    {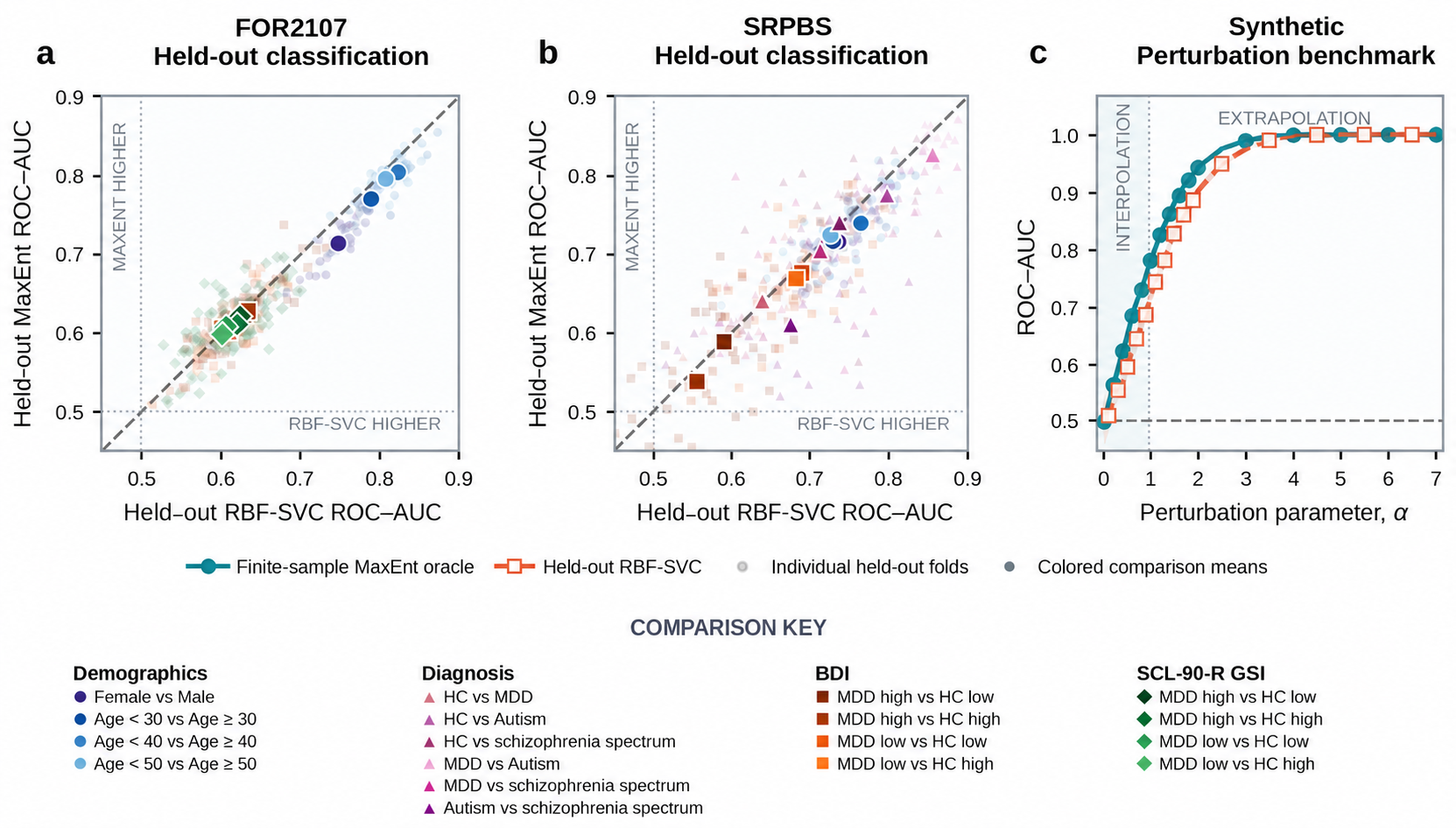}
    \caption{\textbf{Distributional separation and classification across
    FOR2107, controlled simulations, and SRPBS.}
    \textbf{A,} Held-out maximum-entropy and RBF support-vector
    classifier AUCs for 13 comparisons in 1,952 FOR2107 participants,
    using the same 120-dimensional binary connectivity representation,
    with the models trained separately in each cross-validation fold.
    \textbf{B,} The corresponding SRPBS analysis, with the RBF
    support-vector classifier applied to the same binary states,
    across 14 demographic, diagnostic, and BDI comparisons.
    \textbf{C,} Simulated samples from known maximum-entropy
    distributions as the separation parameter $\alpha$ varies. The
    likelihood ratio of the two distributions defines the optimal ranking;
    the best ranking calculated from each simulated sample (the
    finite-sample maximum-entropy oracle) and the held-out RBF-SVC
    both increase with source separation and approach an AUC of one. Curves
    show replicate means and shaded bands show 95\% Student's $t$
    intervals across five generated datasets per $\alpha$.
    In panels A and B, large colored symbols show comparison means and
    translucent symbols show the 25 held-out folds from five repeats of
    five-fold cross-validation; the dashed diagonal denotes equality
    between methods. Both classification panels use the same axis range;
    estimates from individual folds outside the displayed range are not shown.
    Colors distinguish individual comparisons, while marker shapes
    distinguish demographic, diagnostic, BDI, and SCL-90-R GSI
    comparison types. The FOR2107 and SRPBS panels contain different
    sets of clinical comparisons, as indicated by the shared comparison
    key. HC, healthy controls; MDD, major depressive disorder; ASD,
    autism spectrum disorder; SSD, schizophrenia-spectrum disorder; BDI,
    Beck Depression Inventory; SCL-90-R GSI, Symptom
    Checklist-90-Revised Global Severity Index; RBF-SVC,
    radial-basis-function support-vector classifier; AUC, area under the
    receiver-operating-characteristic curve.}
    \label{fig:main_results}
\end{figure}

Across fourteen comparisons, mean AUC was $0.691$ for maximum entropy and $0.706$ for the RBF support-vector classifier applied to the same binary states. The methods' AUCs were strongly correlated ($r=0.975$; Fig.~\ref{fig:main_results}B). Maximum-entropy AUCs ranged from $0.717$ to $0.740$ for the demographic comparisons. HC versus MDD reached $0.642$, and the four BDI comparisons ranged from $0.539$ to $0.676$. The other diagnostic comparisons varied more widely ($0.611$--$0.827$), highest for MDD versus ASD. As in FOR2107, demographic contrasts were more separable than HC versus MDD and the symptom-stratified HC–MDD comparisons. Other diagnostic contrasts were more variable, with some exceeding demographic AUCs. Replication concerned close agreement between maximum-entropy and support-vector classifier estimates across comparisons.

Diagnostic-group distributions across SRPBS sites and acquisition protocols limit AUC interpretation, particularly for ASD and SSD. These values may reflect site and protocol differences alongside diagnosis. The main result is close maximum-entropy and support-vector classifier AUC agreement for identical comparisons and binary brain-state data. Full comparison definitions and uncertainty intervals appear in Appendix~\ref{app:srpbs_results}.

\subsection{Known source distributions determine attainable classification in synthetic data}
\label{sec:results_synthetic}

We next examined classification under known generating distributions. Fixed healthy-control and MDD maximum-entropy models defined a path through their natural parameters: $\alpha=0$ reproduced the reference source, $\alpha=1$ the fitted MDD source, and larger values extrapolated along the same deformation direction. Samples were drawn directly from these tree-structured distributions, whose likelihood ratio supplied the Bayes-optimal score (Appendix~\ref{app:synthetic_methods}).

Classification increased monotonically along this path (Fig.~\ref{fig:main_results}C). At $\alpha=0$, the distributions coincided and both the finite-sample oracle and held-out RBF classifier performed near chance. At $\alpha=1$, finite-sample oracle AUC was $0.781$ and classifier AUC was $0.720$. The classifier followed this trajectory below the oracle at weak and intermediate separations; both approached one as the sources became nearly disjoint. Separately fitted Marburg and M\"unster source models reproduced this pattern (Appendix~\ref{app:site_replication}; Appendix Fig.~\ref{fig:site_replication}b,d).

The synthetic endpoint concerns model-generated samples, not held-out human observations; its AUC need not equal the empirical MDD AUC. The experiment tests whether, under known source distributions, the energy-difference score supplies the population-optimal ranking and nonlinear classification approaches it as separation increases. The remaining gap reflects finite-sample learning and classifier approximation of that ranking.

\subsection{A physical landscape links deformation, work, and snapshot classification}
\label{sec:results_mechanical}

\begin{figure}[H]
    \centering
    \includegraphics[width=\textwidth]
    {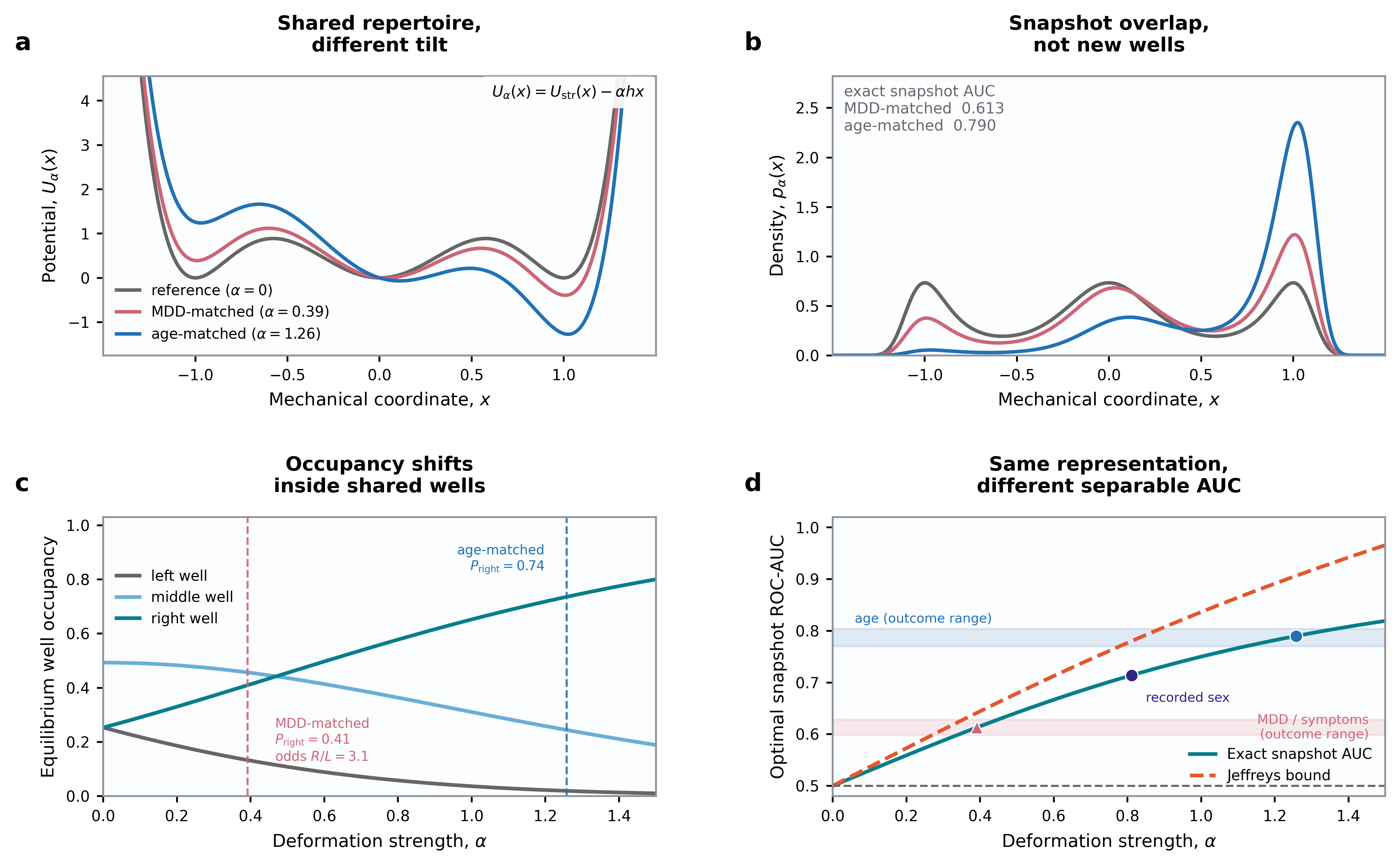}
    \caption{\textbf{A shared physical repertoire can change
    occupancy substantially while remaining weakly classifiable.}
    \textbf{A,} An external field tilts a fixed triple-well
    potential while preserving all three wells. Rose and blue
    mark tilts selected by matching optimal snapshot AUC to the
    empirical MDD value and a value in the empirical age range.
    \textbf{B,} Equilibrium distributions overlap strongly at
    the weaker tilt and separate more at the stronger tilt.
    \textbf{C,} At the weaker tilt, favored-well occupancy
    increases from $0.25$ to $0.41$, with right-to-left odds
    of $3.1{:}1$, while AUC remains approximately $0.61$.
    \textbf{D,} Optimal snapshot AUC increases with deformation
    and remains below the Jeffreys bound. Empirical AUCs
    (MDD $0.613$, recorded sex $0.714$, age approximately
    $0.79$) provide comparison landmarks rather than
    independently predicted outcomes. Mechanical deformation
    is not parameter-matched to the synthetic path.
    Assumptions are one coordinate, a common inverse
    temperature $\beta$, and independent snapshots from
    equilibrated ensembles. The physical interpretation
    applies to this specified system and switching protocol;
    it does not identify neuroimaging statistical energy
    with brain metabolism.}
    \label{fig:mechanical_bridge}
\end{figure}

The mechanical system jointly examined physical and statistical links. An external field tilted the shared triple-well potential,
\begin{equation}
U_\alpha(x)=a x^2(x^2-1)^2-\alpha hx,
\label{eq:main_triple_well}
\end{equation}
preserving all three wells throughout the displayed range (Fig.~\ref{fig:mechanical_bridge}A). Increasing tilt redistributed equilibrium occupancy and progressively separated Boltzmann distributions (Fig.~\ref{fig:mechanical_bridge}B). The specified potential and generating densities permitted joint evaluation of occupancy, distributional divergence, physical work, and optimal snapshot AUC.

Modest deformation substantially changed occupancy while classification remained weak. At $\alpha=0.39$, the favored-well probability increased from $0.25$ to $0.41$, with right-to-left occupancy odds of $3.1{:}1$, yet optimal snapshot AUC was only $0.613$ (Fig.~\ref{fig:mechanical_bridge}C). At $\alpha=1.26$, occupancy concentrated in the favored well ($P_{\mathrm{right}}=0.74$) and AUC reached $0.790$. These operating points matched the empirical MDD AUC and an AUC in the empirical age range; the recorded-sex AUC lay between them. They provide performance landmarks, not fitted biological parameters or independent replication of the empirical hierarchy. Mechanical and synthetic values of $\alpha$ are not interchangeable.

Optimal AUC increased monotonically with deformation and remained below the Jeffreys bound (Fig.~\ref{fig:mechanical_bridge}D). The identity relating Jeffreys divergence to reciprocal dissipated switching work held to numerical precision (Appendix Eq.~\eqref{eq:app_reciprocal_work}). Under the specified equilibrium and switching assumptions, physical landscape deformation connected to occupancy, work, distributional separation, and attainable classification.

This construction demonstrates redistribution within an unchanged repertoire alongside weak snapshot discrimination. It physically realizes the proposed chain but leaves its biological implementation open: the wells are not clinical states, and switching work does not measure metabolic expenditure maintaining neural activity.

\subsection{Existing human evidence does not yet connect metabolic differences to classification limits}
\label{sec:results_evidence}

Finally, the targeted evidence map assessed whether published human
studies could test the proposed metabolic link
\cite{kheirkhah2026targeted}. Coverage was scored on a grid of nine
conditions by five measurement classes---FDG-PET glucose metabolism,
$^{31}$P-MRS ATP/phosphocreatine, quantitative ASL cerebral blood flow,
CMRO$_2$, and ketone/lactate dynamics---giving 45 cells. Only one cell,
schizophrenia FDG-PET, supplied a usable pooled metabolic effect
(Appendix~\ref{app:evidence_coverage}). Matching metabolic and
classification records produced thirteen candidate pairs, but none were
commensurable enough for a correlation. Within pairs, six contrasted different populations or illness
stages. Across pairs, the thirteen metabolic halves fell into six
quantity classes---global absolute glucose metabolism, regional absolute
glucose metabolism, regional relative glucose metabolism, creatine-kinase
rate constants, CMRO$_2$, and mixed units---with no class holding more
than three pairs and no common scale on which diagnoses could be ordered.
No metabolism--AUC correlation was therefore computed
(Appendix~\ref{app:evidence_pairing}).

Reported metabolic effects depended strongly on spatial scope, normalization, measured quantity, modality, and sample definition (Appendix~\ref{app:evidence_effects}; Fig.~\ref{fig:metabolic_evidence_limits}). Demonstrating measurement non-equivalence rather than a pooled estimate, this analysis supports a qualitative conclusion: aggregate evidence lacks commensurate measurements to test the proposed metabolic constraint. This identifies the missing empirical link without supporting or refuting the hypothesis.

\section{Discussion}
\label{sec:discussion}

This paper addresses the recurring disproportion between large behavioral and clinical differences and small, extensively overlapping neuroimaging effects \citep{winter2022quantifying,winter2024systematic}. Dynamical-systems accounts offer an explanation \citep{durstewitz2023reconstructing,scheffer2024dynamical,kheirkhah2026reengineering}. If psychiatric differences arise through changes in occupancy, persistence, and transitions within a largely shared repertoire, a cross-sectional representation may change only modestly even when experience and behavior change substantially \citep{deco2011emerging,breakspear2017dynamic}. The question extends beyond where brain activity is measured to the dynamics generating and maintaining its distribution.

We hypothesize that internal and external inputs deform this activity landscape under joint connectomic and physical resource constraints. These constraints may restrict brain-generated and sustained distributions, consequently limiting discrimination from fixed measurements. This framework connects that biological proposal to an established statistical result: class-conditional distributions determine optimal classification performance \citep{neyman1933ix,devroye1996probabilistic}. It offers a possible physical explanation for weak separability despite clinical consequences.

Empirical, synthetic, and mechanical analyses examine complementary links. In FOR2107, generative and discriminative readouts recovered almost identical hierarchies: age and recorded sex were substantially more distinguishable than diagnosis or symptoms. This pattern replicated separately in Marburg and M\"unster. In the smaller SRPBS sample spanning additional diagnoses, generative and discriminative AUCs again agreed closely across comparisons. Demographic contrasts exceeded HC versus MDD and symptom-stratified HC–MDD separability, but some other diagnostic contrasts exceeded demographic separability. SRPBS therefore supports agreement between methods across broader comparisons, not a uniform demographic-over-diagnostic hierarchy. Uneven diagnostic distributions across SRPBS sites preclude interpreting diagnostic contrasts as pure diagnostic effects. In simulations, known source distributions defined the classification optimum, which nonlinear classifiers approached as separation increased. In the mechanical system, a modest deformation redistributed occupancy while snapshot classification remained weak. Together, these findings suggest classification plateaus reflect limited measured distributional separation alongside estimation and algorithmic limitations. The mechanical example also establishes a work--distribution--classification connection under explicit physical assumptions; its AUC-matched landmarks illustrate compatibility with the empirical pattern rather than independently predicting it.

Biological interpretation requires distinguishing statistical from physical energy. Maximum-entropy energy expresses negative log probability, not glucose use, ATP turnover, oxygen consumption, or heat \citep{jaynes1957information,ezaki2017energy}. Its correspondence to inverse-temperature-scaled physical energy follows from the mechanical system's equilibrium assumptions and cannot simply be transferred to the open, driven brain \citep{jamalabadi2026metabolic}. Nevertheless, neural distributions are physically implemented: signaling and circuit modification require energy, while architecture constrains how activity propagates \citep{attwell2001energy,suarez2020linking,raichle2006brain}. This motivates testing whether physical resources constrain accessible landscape changes' extent, rate, and persistence.

Such a constraint need not act uniformly. Metabolic demands differ across neural processes and circuits \citep{jamadar2025metabolic}, while peripheral and interoceptive signals couple neural resource availability to organismal physiology \citep{figlewicz2009insulin,scherer2021brain,mehrhof2025interoceptive}. The budget is therefore distributed across an embodied system. Analysis in Appendix~\ref{app:resource_bound} formalizes this by combining available resources, directional deformation costs, and their effects on the measured distribution. Crucially, those costs must be calibrated independently of distributional divergence. The bound is testable but does not estimate FOR2107 or SRPBS metabolic expenditure.

\begin{figure}[H]
    \centering
    \includegraphics[width=1.0\linewidth,keepaspectratio]
    {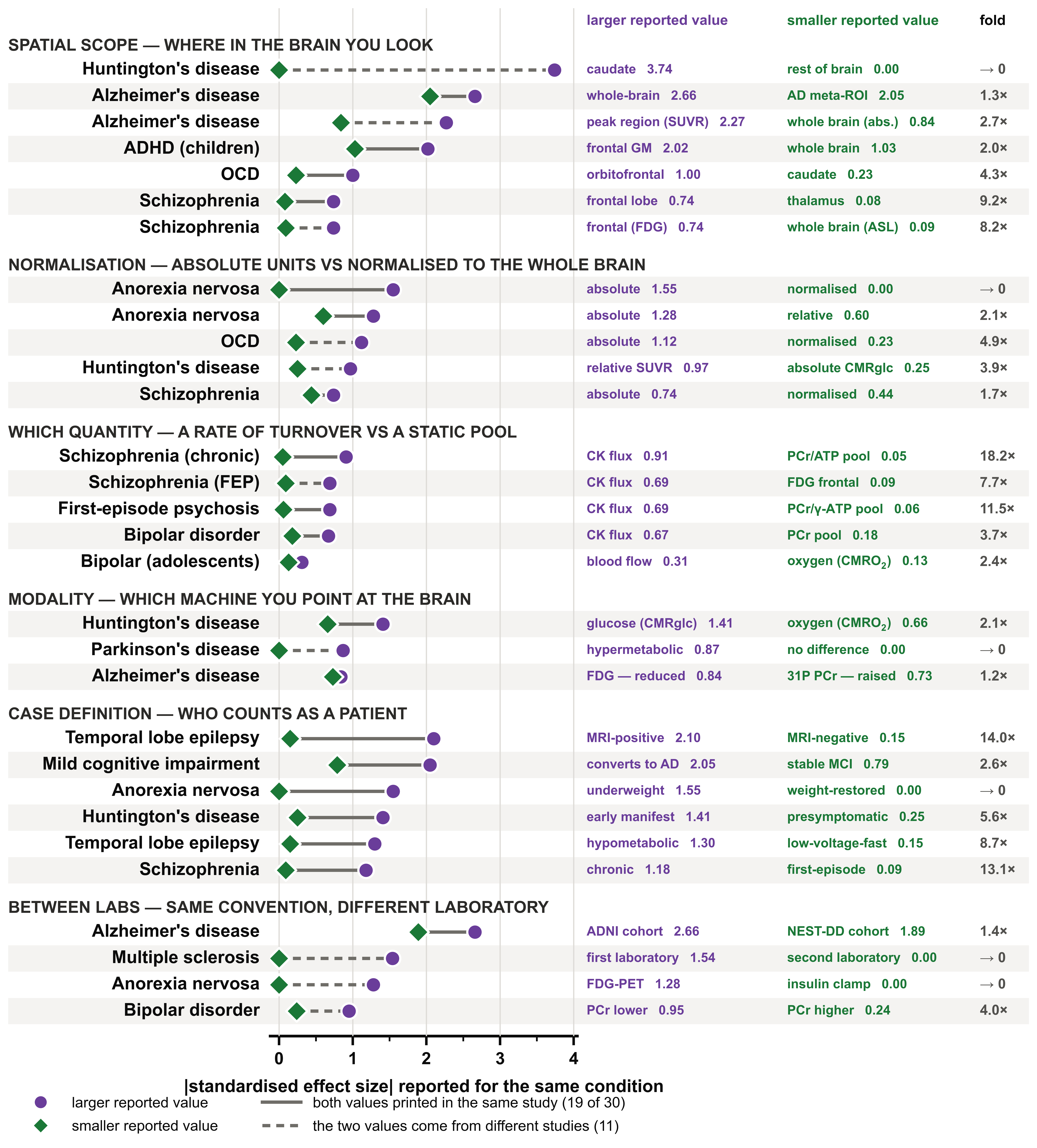}
    \caption{\textbf{Reported brain-metabolic effect sizes
    are strongly measurement- and scale-dependent.}
    Illustrative paired contrasts show how absolute
    standardized metabolic effects vary with spatial scope,
    normalization, measured quantity, modality, case definition,
    and laboratory. Circles show larger and diamonds smaller
    effects; solid lines identify within-study comparisons
    and dashed lines between-study comparisons. Across
    30 contrasts, 19 were obtained within the same study
    and six included a zero estimate. Among pairs with two nonzero estimates, the median larger-to-smaller ratio was $3.9$ (range $1.2$--$18.2$). Three pairs contained
    effects with opposite directions. These exploratory
    contrasts illustrate measurement non-equivalence;
    they are not pooled estimates or evidence that every
    condition has a fourfold uncertainty. Sources, identifiers, and sample sizes for every contrast are given in Supplementary Table~S5, which also marks the three pairs with effects of opposite direction (Appendix~\ref{app:evidence_effects}).}
    \label{fig:metabolic_evidence_limits}
\end{figure}

Human evidence could not supply this connection: the targeted map found no commensurable candidate
metabolism--classification pairs: within pairs the halves often describe
different populations or stages, and across pairs the metabolic halves
share no common scale (Appendix~\ref{app:evidence_map}). Reported metabolic effects also depended on spatial scope, normalization, modality, and population definition (Fig.~\ref{fig:metabolic_evidence_limits}). These findings identify a measurement gap rather than evidence against the hypothesis. Establishing this link requires measuring input, structure, calibrated energy use, activity distributions, and behavior in the same individuals at compatible scales.

A focused experimental program could use repeated measurements to relate metabolic status to occupancy, dwell times, transitions, and standardized perturbation responses \citep{poldrack2017precision,durstewitz2023reconstructing}. Controlled changes in resource availability would provide a stronger test than cross-sectional associations, provided their effects on arousal, vascular responses, and neural signaling are distinguished. The central prediction concerns the range of distributional changes accessible under measured resource constraints. Reliable violations would challenge a specified quantitative bound after checking its assumptions and calibration. An absent simple expenditure--separation correlation would be less decisive: systems need not operate at their resource limit. Behavioral measurements would establish whether constrained changes affect symptoms and function.

Several limitations remain. Diagnostic heterogeneity, measurement noise, temporal averaging, and discarded information can also produce overlap. Empirical maximum-entropy references depend on the binary representation and sparse model; more informative features may improve discrimination. Between-person connectivity distributions also do not directly establish within-person attractors or a shared dynamical landscape \citep{lurie2020questions}. These results neither identify metabolic constraints as causing weak classification nor exclude informative subgroups or longitudinal signatures. They establish a distributional argument and physically realize the proposed connection.

For psychiatry, substantial dysfunction may emerge within a largely shared repertoire. Progress requires measuring repertoire organization, change, and which changes the organism can sustain. The framework formalizes this possibility and identifies human measurements needed to determine whether physiological constraints contribute to persistent neuroimaging classification limits.

\section{Conclusion}
\label{sec:conclusion}

Profound clinical differences need not produce sharply distinguishable brain images. Our findings support a distributional explanation for weak classification and motivate a broader hypothesis: psychiatric disorders deform a largely shared neural repertoire within structural and metabolic constraints. The human analyses establish agreement in measured separability; the mechanical system demonstrates how occupancy can change while snapshot discrimination remains weak. Testing the proposed physiological constraint requires joint measurements of brain activity, architecture, and metabolism. Progress depends on understanding which neural changes produce clinical consequences and how much of that change our measurements can reveal.

\section{Declarations}

\subsection{Ethical Approval}
All participants provided written informed consent, and the study was approved by the ethics committees of the Medical Faculties at the University of Marburg (AZ: 07/14) and University of M\"unster (AZ: 2014-422-b-S). Participants received financial compensation.
For the SRPBS dataset, all participants provided written informed consent, and all recruitment procedures and experimental protocols were approved by the institutional review boards of the principal investigators' respective institutions \citep{tanaka2021multi}. The SRPBS data were obtained in anonymized form from the DecNef Project Brain Data Repository (\href{https://bicr-resource.atr.jp/srpbsopen/}{https://bicr-resource.atr.jp/srpbsopen/}) and used in accordance with that repository's terms of use \citep{tanaka2021multi}.

\subsection{FOR2107 Consortium}
Frederike Stein, Florian Thomas-Odenthal, Paula Usemann, Lea Teutenberg, Benjamin Straube, Nina Alexander, Andreas Jansen, Nooshin Javaheripour, Susanne Meinert, Elisabeth J. Leehr, Tiana Borgers, Kira Flinkenfl\"ugel, Dominik Grotegerd, Lukas Fisch, Julia Hubbert, Christoph Jurischka, Judith Krieger, Anne Hildebrand, Roman Vogler, Sabrina Teckentrup, Sara Papenbrock, Franziska Herrmann, Elilarasi Ratnalingam, Theresa Slump, Hannah Meinert.

\subsection{Funding Acknowledgment}
This work was supported in part by the German Research Foundation (DFG) through the Collaborative Research Center SFB/TRR 393 (project grant no.\ 521379614), by a research grant from the von Behring-R\"ontgen Stiftung (grant no.\ 70\_00038), by a research grant from the University Hospital of Gie{\ss}en and Marburg (UKGM, grant no.\ 1/2024 MR), and by the Excellence Cluster EXC3066 ``The Adaptive Mind''. This paper is further partially supported by ERA-NET NEURON JTC 2024 (BRAWO project, grant no.\ ER-2024-23684536). SGH receives financial support by the Alexander von Humboldt Foundation (as part of the Alexander von Humboldt Professur), the Hessisches Ministerium f\"ur Wissenschaft und Kunst (as part of the LOEWE Spitzenprofessur) and the DYNAMIC center, funded by the LOEWE program of the Hessian Ministry of Science and Arts (Grant Number: LOEWE1/16/519/03/09.001(0009)/98).
He also receives compensation for his work as editor from the American Psychological Association and royalties and payments for his work from various publishers. UD was funded by the German Research Foundation (DFG, grant FOR2107 DA1151/5-1, DA1151/5-2, DA1151/9-1, DA1151/10-1, DA1151/11-1 to UD; SFB/TRR 393, project grant no 521379614) and the Interdisciplinary Center for Clinical Research (IZKF) of the medical faculty of M\"unster (grant Dan3/016/26 to UD). The funders had no involvement in the study design, data collection, management, analysis, interpretation, report writing, or the decision to submit the report for publication. The authors retained full responsibility for the decision to submit for publication.

\subsection{Data and Code Availability}
The FOR2107 data are part of the Marburg/M\"unster Affective Disorder cohort and can be accessed by requesting permission from Prof.\ Tilo Kircher via the Collaborative Research Center (CRC) \href{https://www.sfb-trr393.de/}{https://www.sfb-trr393.de/}.
The SRPBS analysis used the unrestricted SRPBS Multi-disorder MRI Dataset, obtained from the DecNef Project Brain Data Repository (\href{https://bicr-resource.atr.jp/srpbsopen/}{https://bicr-resource.atr.jp/srpbsopen/}) \citep{tanaka2021multi}. Data used in the preparation of this work were obtained from the DecNef Project Brain Data Repository (\href{https://bicr-resource.atr.jp/srpbsopen/}{https://bicr-resource.atr.jp/srpbsopen/}), gathered by a consortium as part of the Japanese Strategic Research Program for the Promotion of Brain Science (SRPBS) supported by the Japanese Advanced Research and Development Programs for Medical Innovation (AMED). Access requires registration and agreement to the repository terms; the data may not be copied or redistributed. This is the SRPBS imaging release hosted by the DecNef repository, not the separate DecNef decoded-neurofeedback collection.
All code used to estimate the parameters and generate the results presented in this paper will be made publicly available upon acceptance of the manuscript at: \href{https://www.psycontrol-lab.de/}{https://www.psycontrol-lab.de/}.

\subsection{Conflict of Interest}
The authors declare no competing interests.

\subsection{Author Contributions}
Conceptualization, Methodology, and Validation: TH, HJ.
Statistical Analysis: FK, HJ, SB, MK, BS.
Data Acquisition: MK, BS, MS, UD, TK, IN, TH, HJ.
Writing -- Original Draft: FK, HJ, MK, MS.
Writing -- Review and Editing: all authors.

\newpage

\bibliography{sn-bibliography}

\newpage
\begin{appendices}

The appendices provide the mathematical derivations underlying Section~\ref{sec:theoretical_framework}, the empirical and synthetic methods supporting Section~\ref{sec:results}, supplementary results, and the targeted evidence map reported in Section~\ref{sec:results_evidence} and discussed in Section~\ref{sec:discussion}.

\section{Mathematical derivations}
\label{app:mathematics}

\subsection{Notation and scope}

This appendix supplies the derivations used in Section~\ref{sec:theoretical_framework} and the exact mechanical analysis in Section~\ref{sec:results_mechanical}. Conditions are indexed by $g\in\{0,1\}$. In the mechanical example, $U_g(x)$ denotes physical potential energy, $\beta=(k_{\mathrm B}T)^{-1}$ is inverse temperature, and $p_g(x)$ is the associated equilibrium density. In the maximum-entropy analyses, $E_g(\boldsymbol z)$ denotes dimensionless statistical energy and $p_g(\boldsymbol z)$ is a fitted probability mass function over binary latent states. These uses of ``energy'' become mathematically identical for an equilibrium Boltzmann system after multiplication of physical energy by $\beta$. For fMRI, however, $E_g$ is a negative-log-probability representation and is not a direct measurement of ATP turnover, glucose use, oxygen consumption, or dissipated heat. The mechanical derivation therefore establishes the deformation--work--distribution--classification chain under explicit equilibrium assumptions; the empirical analysis tests its distributional and classification consequences only.

\subsection{From mechanical to statistical energy}
\label{app:mechanical}

Consider a one-dimensional coordinate $x$, velocity $v$, mass $m$, and condition-specific potential $U_g(x)$. The Hamiltonian is
\begin{equation}
H_g(x,v)=\frac{1}{2}mv^2+U_g(x).
\label{eq:app_hamiltonian}
\end{equation}
At thermal equilibrium, the canonical phase-space density is
\begin{equation}
\rho_g(x,v)=\frac{\exp[-\beta H_g(x,v)]}{\mathcal Z_g}.
\end{equation}
Integrating out velocity yields
\begin{equation}
p_g(x)=\int \rho_g(x,v)\,dv
=\frac{\exp[-\beta U_g(x)]}{Z_g},
\qquad
Z_g=\int \exp[-\beta U_g(x)]\,dx.
\label{eq:app_boltzmann}
\end{equation}
Thus, in the energy-based convention $p_g(x)=Z_g^{-1}\exp[-E_g(x)]$, the dimensionless statistical energy is $E_g(x)=\beta U_g(x)$ up to an additive constant. The same stationary density follows from overdamped Langevin dynamics
\begin{equation}
dX_t=-\mu\,\partial_xU_g(X_t)\,dt
+\sqrt{2\mu k_{\mathrm B}T}\,dW_t,
\end{equation}
provided detailed balance holds.

\subsection{Energy differences and Bayes-optimal ranking}
\label{app:optimal_ranking}

\noindent\textbf{Lemma A1 (energy--likelihood correspondence).}
For two Boltzmann densities at a common inverse temperature, ranking observations by $U_0(x)-U_1(x)$ is equivalent to ranking them by the likelihood ratio $p_1(x)/p_0(x)$.

\medskip
\noindent\emph{Proof.}
From Eq.~\eqref{eq:app_boltzmann},
\begin{equation}
\ell(x)=\log\frac{p_1(x)}{p_0(x)}
=\beta[U_0(x)-U_1(x)]+\log\frac{Z_0}{Z_1}.
\label{eq:app_llr_mechanical}
\end{equation}
The partition-function term is constant across observations and $\beta>0$; consequently, both scores induce the same ordering. The Neyman--Pearson lemma then implies that thresholding this ordering gives the uniformly most powerful test at every false-positive rate and hence the optimal receiver-operating-characteristic curve \citep{neyman1933ix}. \hfill$\square$

For independent $X_1\sim p_1$ and $X_0\sim p_0$, the corresponding AUC is the weighted Mann--Whitney probability
\begin{equation}
\AUC^*
=\Prob[\ell(X_1)>\ell(X_0)]
+\frac{1}{2}\Prob[\ell(X_1)=\ell(X_0)].
\label{eq:app_auc_pairwise}
\end{equation}
The tie probability is zero when the likelihood-ratio score has a continuous distribution; continuous observations alone do not guarantee this. For discrete maximum-entropy states it is retained explicitly.

The same result applies directly to an energy-based model over a discrete representation $\boldsymbol z$,
\begin{equation}
p_g(\boldsymbol z)=\frac{\exp[-E_g(\boldsymbol z)]}{Z_g}.
\end{equation}
Its log likelihood ratio is
\begin{equation}
\log\frac{p_1(\boldsymbol z)}{p_0(\boldsymbol z)}
=E_0(\boldsymbol z)-E_1(\boldsymbol z)+\log\frac{Z_0}{Z_1}.
\end{equation}
Because the partition-function term is constant across observations, the energy difference $q(\boldsymbol z)=E_0(\boldsymbol z)-E_1(\boldsymbol z)$ produces the optimal ordering. For independent $\boldsymbol Z_1\sim p_1$ and $\boldsymbol Z_0\sim p_0$, the corresponding maximum-entropy AUC is therefore
\begin{equation}
\mathrm{AUC}_{\mathrm{ME}}
=
\Pr\left[q(\boldsymbol Z_1)>q(\boldsymbol Z_0)\right]
+\frac{1}{2}\Pr\left[q(\boldsymbol Z_1)=q(\boldsymbol Z_0)\right].
\label{eq:app_maxent_auc}
\end{equation}
No classifier operating on the same representation can improve on this population ordering when the two distributions are known exactly \citep{neyman1933ix,cover2006elements}. With fitted empirical distributions, the value remains conditional on the representation and model family, as detailed in Appendices~\ref{app:maxent_methods} and \ref{app:conditional_reference}.

\subsection{Information-theoretic bound on optimal AUC}
\label{app:auc_bound}

Define the Jeffreys divergence
\begin{equation}
J(p_0,p_1)=\KL(p_1\Vert p_0)+\KL(p_0\Vert p_1).
\end{equation}

\noindent\textbf{Theorem A1 (Jeffreys bound on likelihood-ratio AUC).}
For any mutually absolutely continuous $p_0$ and $p_1$,
\begin{equation}
\AUC^*\leq
\min\left\{1,\frac{1}{2}+\sqrt{\frac{J(p_0,p_1)}{8}}\right\}.
\label{eq:app_auc_bound}
\end{equation}

\medskip
\noindent\emph{Proof.}
Let $R=p_1\times p_0$ and $S=p_0\times p_1$ be distributions over ordered pairs. Define the event $A=\{(x,y):\ell(x)>\ell(y)\}$ and break ties randomly. Swapping the two coordinates gives
\begin{equation}
R(A)-S(A)=2\AUC^*-1.
\end{equation}
By the variational definition of total variation and Pinsker's inequality,
\begin{equation}
2\AUC^*-1
\leq \TV(R,S)
\leq\sqrt{\frac{\KL(R\Vert S)}{2}}.
\end{equation}
Product additivity gives
\begin{align}
\KL(R\Vert S)
&=\KL(p_1\times p_0\Vert p_0\times p_1)\\
&=\KL(p_1\Vert p_0)+\KL(p_0\Vert p_1)
=J(p_0,p_1),
\end{align}
which yields Eq.~\eqref{eq:app_auc_bound}. \hfill$\square$

The inequality is distribution-free but generally not tight. The likelihood-ratio AUC in Eq.~\eqref{eq:app_auc_pairwise}, rather than the Jeffreys inequality, defines the population optimum for known source distributions. For human observations, the fitted score supplies the conditional reference described in Appendix~\ref{app:conditional_reference}. Equality would require both that the likelihood-ordering event attain the total-variation supremum and that the Pinsker step be tight.

\subsection{Jeffreys divergence and reciprocal dissipated work}
\label{app:reciprocal_work}

The equilibrium free energy is $F_g=-\beta^{-1}\log Z_g$. Consider an instantaneous switch from $U_0$ to $U_1$ while the initial state remains distributed according to $p_0$. Its mean dissipated work is
\begin{equation}
W_{\mathrm{diss}}^{0\rightarrow1}
=\E_0[U_1-U_0]-(F_1-F_0).
\label{eq:app_forward_work}
\end{equation}
Direct substitution of the Boltzmann densities gives
\begin{align}
\KL(p_0\Vert p_1)
&=\E_0\left[\log\frac{p_0(X)}{p_1(X)}\right]\\
&=\beta\E_0[U_1-U_0]+\log\frac{Z_1}{Z_0}\\
&=\beta W_{\mathrm{diss}}^{0\rightarrow1}.
\end{align}
The reverse switch similarly satisfies
\begin{equation}
\KL(p_1\Vert p_0)=\beta W_{\mathrm{diss}}^{1\rightarrow0}.
\end{equation}
Therefore,
\begin{equation}
J(p_0,p_1)
=\beta\left(W_{\mathrm{diss}}^{0\rightarrow1}
+W_{\mathrm{diss}}^{1\rightarrow0}\right).
\label{eq:app_reciprocal_work}
\end{equation}
This identity applies to the stated reciprocal switching protocol between equilibrium systems \citep{kawai2007dissipation}. Combining Eqs.~\eqref{eq:app_auc_bound} and \eqref{eq:app_reciprocal_work} gives the physical classification bound underlying Eqs.~\eqref{eq:framework_bounds} and \eqref{eq:framework_work}:
\begin{equation}
\AUC^*\leq
\min\left\{1,
\frac{1}{2}+\sqrt{\frac{\beta
(W_{\mathrm{diss}}^{0\rightarrow1}+W_{\mathrm{diss}}^{1\rightarrow0})}{8}}
\right\}.
\end{equation}

\subsection{Tilted triple-well system}
\label{app:triple_well}

The mechanical illustration in Section~\ref{sec:results_mechanical} (Eqs.~\eqref{eq:framework_mechanical} and \eqref{eq:main_triple_well}) used
\begin{equation}
U_{\alpha}(x)=a x^2(x^2-1)^2-\alpha h x,
\qquad a>0,
\label{eq:app_triple_well}
\end{equation}
with $a=6$, $h=1$, and $\beta=1.5$. Its stationary points satisfy
\begin{equation}
\partial_xU_{\alpha}(x)
=2ax(x^2-1)(3x^2-1)-\alpha h=0,
\end{equation}
and their curvature is
\begin{equation}
\partial_x^2U_{\alpha}(x)=2a(15x^4-12x^2+1).
\end{equation}
Saddle-node loss of multistability occurs when both derivatives vanish. Writing $y=x^2$ gives
\begin{equation}
y_{\pm}=\frac{6\pm\sqrt{21}}{15}.
\end{equation}
For the numerical parameters above, the first positive critical tilt has magnitude approximately $2.393$. The displayed interval $0\leq\alpha\leq1.5$ therefore retains three stable wells.

Relative to $\alpha=0$, the likelihood-ratio score is
\begin{equation}
q_{\alpha}(x)=U_0(x)-U_{\alpha}(x)=\alpha h x.
\end{equation}
For $\alpha>0$ and the specified $h>0$, this score is monotonic in $x$, and the exact AUC becomes the one-dimensional integral
\begin{equation}
\AUC_{\alpha}^*
=\int_{-\infty}^{\infty}p_{\alpha}(x)F_0(x)\,dx,
\qquad
F_0(x)=\int_{-\infty}^{x}p_0(y)\,dy.
\label{eq:app_triple_auc}
\end{equation}
The Jeffreys divergence simplifies to
\begin{equation}
J_{\alpha}
=\beta\alpha h
\left(\langle x\rangle_{\alpha}-\langle x\rangle_0\right).
\label{eq:app_triple_J}
\end{equation}
Because
\begin{equation}
\frac{d\langle x\rangle_{\alpha}}{d\alpha}
=\beta h\operatorname{Var}_{\alpha}(X),
\end{equation}
Eq.~\eqref{eq:app_triple_J} can also be written as
\begin{equation}
J_{\alpha}=\beta^2h^2\alpha
\int_0^{\alpha}\operatorname{Var}_t(X)\,dt.
\end{equation}
The numerical curves in Fig.~\ref{fig:mechanical_bridge} were evaluated by deterministic quadrature on $120{,}001$ equally spaced points over $x\in[-2.25,2.25]$. The maximum absolute discrepancy between Jeffreys divergence and reciprocal dimensionless dissipated work was $6.7\times10^{-16}$.

\subsection{Local Fisher-information limit}

The Fisher information along the deformation path is
\begin{equation}
I(\alpha)
=\E_{\alpha}\left[
\left(\partial_{\alpha}\log p_{\alpha}(X)\right)^2
\right]
=(\beta h)^2\operatorname{Var}_{\alpha}(X).
\end{equation}
Consequently,
\begin{equation}
J_{\alpha}=\alpha\int_0^{\alpha}I(t)\,dt
=\alpha^2I(0)+O(\alpha^3).
\label{eq:app_fisher_local}
\end{equation}
This explains why the Fisher approximation agrees with the exact work-information relation only for small landscape deformations.

For comparison, consider equal-curvature harmonic landscapes
\begin{equation}
U_g(x)=\frac{k}{2}(x-m_g)^2.
\end{equation}
Their equilibrium densities are Gaussian with common variance $\sigma^2=(\beta k)^{-1}$. With $\delta=|m_1-m_0|/\sigma$,
\begin{equation}
J=\delta^2,
\qquad
\AUC^*=\Phi\left(\frac{\delta}{\sqrt{2}}\right)
=\Phi\left(\sqrt{\frac{J}{2}}\right),
\label{eq:app_harmonic_auc}
\end{equation}
where $\Phi$ is the standard-normal cumulative distribution function. Equation~\eqref{eq:app_harmonic_auc} provides an analytically tractable Gaussian comparison with the triple-well system.

\subsection{Exponential-family and maximum-entropy generalization}

Let
\begin{equation}
p_{\boldsymbol\theta}(\boldsymbol z)
=\exp\left[
\boldsymbol\theta^{\top}\boldsymbol T(\boldsymbol z)
-A(\boldsymbol\theta)
\right]
\end{equation}
be an exponential family with sufficient statistics $\boldsymbol T$ and log-partition function $A$ \citep{wainwright2008graphical}. Along the natural-parameter path
\begin{equation}
\boldsymbol\theta_{\alpha}
=\boldsymbol\theta_0+\alpha\Delta\boldsymbol\theta,
\qquad
\Delta\boldsymbol\theta=\boldsymbol\theta_1-\boldsymbol\theta_0,
\label{eq:app_theta_path}
\end{equation}
the score between $p_{\boldsymbol\theta_0}$ and $p_{\boldsymbol\theta_{\alpha}}$ is, up to a constant,
\begin{equation}
\ell_{\alpha}(\boldsymbol z)
=\alpha\Delta\boldsymbol\theta^{\top}\boldsymbol T(\boldsymbol z)
-[A(\boldsymbol\theta_{\alpha})-A(\boldsymbol\theta_0)].
\end{equation}
The Jeffreys divergence is
\begin{align}
J_{\alpha}
&=\alpha\Delta\boldsymbol\theta^{\top}
(\boldsymbol\mu_{\alpha}-\boldsymbol\mu_0)\\
&=\alpha\int_0^{\alpha}
\Delta\boldsymbol\theta^{\top}
\boldsymbol I(\boldsymbol\theta_t)
\Delta\boldsymbol\theta\,dt,
\label{eq:app_exp_family_J}
\end{align}
where $\boldsymbol\mu_{\alpha}=\E_{\alpha}[\boldsymbol T(\boldsymbol Z)]$ and $\boldsymbol I(\boldsymbol\theta)=\nabla^2A(\boldsymbol\theta)$ is the Fisher-information matrix. For a pairwise binary maximum-entropy model, $\boldsymbol T$ contains node states and selected pairwise products \citep{ezaki2017energy,wainwright2008graphical}. Equation~\eqref{eq:app_exp_family_J} is the statistical counterpart of the triple-well deformation path and supplies the interpolation used in the synthetic analysis.

\subsection{Local resource-constrained separability}
\label{app:resource_bound}

Let $\delta\boldsymbol\theta$ be a sufficiently small displacement in the parameter space of a regular statistical model. The local expansion of Jeffreys divergence is
\begin{equation}
J(p_{\boldsymbol\theta},p_{\boldsymbol\theta+\delta\boldsymbol\theta})
=
\delta\boldsymbol\theta^{\top}
\boldsymbol I(\boldsymbol\theta)
\delta\boldsymbol\theta
+O(\lVert\delta\boldsymbol\theta\rVert^3).
\label{eq:app_local_jeffreys}
\end{equation}
Suppose that admissible displacements additionally satisfy the independently specified quadratic accessibility constraint
\begin{equation}
\delta\boldsymbol\theta^{\top}
\boldsymbol C
\delta\boldsymbol\theta
\leq B,
\qquad \boldsymbol C\succ0.
\label{eq:app_accessibility_constraint}
\end{equation}
Writing $\boldsymbol v=\boldsymbol C^{1/2}\delta\boldsymbol\theta$ gives $\boldsymbol v^{\top}\boldsymbol v\leq B$ and transforms the leading term in Eq.~\eqref{eq:app_local_jeffreys} into
\begin{equation}
\boldsymbol v^{\top}
\left[
\boldsymbol C^{-1/2}
\boldsymbol I(\boldsymbol\theta)
\boldsymbol C^{-1/2}
\right]
\boldsymbol v.
\end{equation}
The Rayleigh-quotient bound therefore yields
\begin{equation}
J(p_{\boldsymbol\theta},p_{\boldsymbol\theta+\delta\boldsymbol\theta})
\leq
B\,\lambda_{\max}\!\left[
\boldsymbol C^{-1/2}
\boldsymbol I(\boldsymbol\theta)
\boldsymbol C^{-1/2}
\right]
+O(B^{3/2}).
\label{eq:app_resource_jeffreys}
\end{equation}
Combining this result with Eq.~\eqref{eq:app_auc_bound} gives the corresponding local classification constraint
\begin{equation}
\AUC^*-\frac{1}{2}
\leq
\sqrt{
\frac{
B\,\lambda_{\max}\!\left[
\boldsymbol C^{-1/2}
\boldsymbol I(\boldsymbol\theta)
\boldsymbol C^{-1/2}
\right]
+O(B^{3/2})
}{8}}.
\label{eq:app_resource_auc}
\end{equation}
If $\boldsymbol C=\boldsymbol W_{\theta}^{-1}$ for a controllability Gramian $\boldsymbol W_{\theta}$ defined in the same parameter space, the matrix in Eqs.~\eqref{eq:app_resource_jeffreys}--\eqref{eq:app_resource_auc} becomes $\boldsymbol W_{\theta}^{1/2}\boldsymbol I(\boldsymbol\theta)\boldsymbol W_{\theta}^{1/2}$. When controllability is initially defined in a neural-state space, an additional local mapping from neural states to distribution parameters is required. This is the local bound summarized in Eq.~\eqref{eq:framework_resource}; the remainder assumes fixed positive-definite $\boldsymbol C$ and a sufficiently small budget. The quantity $B$ represents metabolic energy only when the input cost is independently calibrated in physical units. Thus, this derivation identifies the additional assumptions needed to connect physical accessibility to distributional and classification limits; it is not used to infer metabolic expenditure from the FOR2107 distributions.

\subsection{Why the empirical reference is conditional}
\label{app:conditional_reference}

Let $X$ denote the original functional-connectivity representation and $Z=\phi(X)$ the learned binary representation. The likelihood ratio of the true class-conditional distributions of $Z$ would define the Bayes-optimal AUC available from $Z$. In practice, $p_g(Z)$ is approximated by a tree-structured pairwise model $\widehat p_g$. The reported empirical value is therefore
\begin{equation}
\widehat{\AUC}_{\mathrm{ME}}
=\AUC\!\left(\log\frac{\widehat p_1(Z)}{\widehat p_0(Z)}\right),
\end{equation}
not the unrestricted Bayes AUC of $X$. A classifier may exceed this reference if it uses information discarded by $\phi$, exploits higher-order structure absent from $\widehat p_g$, or benefits from finite-sample fluctuations. In the synthetic experiment, by contrast, $\widehat p_g$ is the data-generating family; the source likelihood ratio is then a genuine Bayes ceiling for observations from that source.

\section{Participants, imaging, and empirical features}
\label{app:empirical_methods}

This section provides the cohort, acquisition, preprocessing, and target definitions underlying the FOR2107 and SRPBS results in Fig.~\ref{fig:main_results}A,B.

\subsection{Participants and ethics}

Participants were recruited through the Marburg--M\"unster Affective Disorders Cohort Study (FOR2107/MACS) from in-patient services, collaborating clinical institutions, and public advertisements \citep{kircher2019neurobiology}. All participants provided written informed consent and received financial compensation. The study was conducted in accordance with the Declaration of Helsinki and approved by the ethics committees of the Medical Faculties of the University of Marburg (AZ 07/14) and the University of M\"unster (AZ 2014-422-b-S).

After exclusion of incomplete records required for the primary analysis, the analytic cohort contained 1,952 participants: 1,129 healthy controls (HC) and 823 participants with major depressive disorder (MDD). Descriptive characteristics are shown in Table~\ref{tab:app_demographics}.

\begin{table}[ht]
\centering
\caption{Demographic characteristics of the analytic cohort.}
\label{tab:app_demographics}
\begin{tabular}{lccc}
\toprule
Variable & MDD ($n=823$) & HC ($n=1{,}129$) & All ($n=1{,}952$)\\
\midrule
Female, \% & 63 & 63 & 63\\
Age, years, mean (SD) & 36.78 (13.38) & 35.42 (13.54) & 35.99 (13.48)\\
Education, years, mean (SD) & 13.31 (2.74) & 13.99 (2.68) & 13.71 (2.73)\\
\bottomrule
\end{tabular}
\end{table}

The second dataset was the publicly available Strategic Research Program for Brain Sciences (SRPBS) Multi-disorder MRI Dataset, which comprises structural and resting-state functional MRI data acquired at multiple sites across Japan. All participants provided written informed consent, and all recruitment procedures and experimental protocols were approved by the institutional review boards of the participating institutions \citep{tanaka2021multi}.

From the 1,410 participants in this dataset, we included the 1,318 healthy controls (HC) and participants diagnosed with major depressive disorder (MDD), schizophrenia-spectrum disorder (SSD), or autism spectrum disorder (ASD); participants with other diagnoses (bipolar disorder, chronic pain, dysthymia, or other conditions; $n=92$) were not considered. After exclusion of 418 participants with missing or incomplete imaging data, the analytic cohort contained 900 participants: 500 HC, 220 participants with MDD, 55 participants with SSD, and 125 participants with ASD. Descriptive characteristics are shown in Table~\ref{tab:app_demographics_srpbs}.

\begin{table}[ht]
\centering
\caption{Demographic characteristics of the SRPBS analytic cohort.}
\label{tab:app_demographics_srpbs}
\begin{tabular}{lccccc}
\toprule
Variable & MDD ($n=220$) & SSD ($n=55$) & ASD ($n=125$) & HC ($n=500$) & All ($n=900$)\\
\midrule
Female, \% & 48 & 29 & 13 & 51 & 44\\
Age, years, mean (SD) & 42.05 (11.98) & 35.40 (11.10) & 32.47 (8.03) & 41.22 (15.19) & 39.85 (13.79)\\
\bottomrule
\end{tabular}
\end{table}

\subsection{MRI acquisition}

Structural and resting-state functional MRI for the Marburg--M\"unster Affective Disorders Cohort Study (FOR2107/MACS)  were acquired at the University Hospitals of Marburg and M\"unster. Marburg used a 3-T Siemens Tim Trio scanner with a 12-channel head coil; M\"unster used a 3-T Siemens Prisma scanner with a 20-channel head coil. T1-weighted MP-RAGE images had 1-mm isotropic resolution, a 256-mm field of view, and 176 sagittal slices in Marburg or 192 sagittal slices in M\"unster. Resting-state fMRI was acquired for 8 minutes while participants rested with their eyes closed.
The SRPBS structural and resting-state functional MRI data were acquired on 3-T Siemens (Verio, Tim Trio, Trio, Spectra) and GE (Signa HDxt, Discovery MR750w) scanners at 11 sites using 13 acquisition protocols, which differed in scanner model and acquisition parameters; full details are reported in \citep{tanaka2021multi}. T1-weighted images had an approximately 1-mm isotropic resolution and a 225--256-mm field of view. Resting-state fMRI was acquired with gradient-echo echo-planar imaging (TR = 2.0--2.7~s, TE = 27--31~ms, flip angle = 80--90\textdegree, in-plane resolution = 3.0--4.0~mm, slice thickness = 3.0--4.0~mm) for approximately 5--10 minutes, while participants were instructed to relax, not to fall asleep, and to keep their eyes open fixating on a black cross.

\subsection{Image preprocessing and functional-connectivity construction}

T1-weighted data underwent bias-field correction with \texttt{fsl\_anat} and brain extraction with SynthStrip \citep{hoopes2022}. Functional data were processed using FSL 6.0.7.6 \citep{smith2004advances}. The preprocessing workflow comprised MCFLIRT motion correction, BET brain extraction, 5-mm full-width-at-half-maximum Gaussian spatial smoothing, and temporal high-pass filtering with a 100-s cutoff period. Functional images were registered to the participant's T1-weighted image and normalized to 2-mm MNI152 space using 12-degree-of-freedom affine registration. Within-subject spatial independent-component analysis was performed with MELODIC using automatic dimensionality estimation, followed by FSL-FIX removal of components classified as motion, physiological, or scanner artifact \citep{salimi2014automatic}. Both datasets were preprocessed using an identical pipeline.

Regional time series were extracted using the Schaefer cortical parcellation with 100 regions assigned to 17 functional networks \citep{schaefer2018local, yeo2011organization}, provided in 2-mm FSL MNI152 space. For each region, the BOLD signal was averaged across all voxels with valid signal, excluding voxels outside the functional field of view, and the resulting time series were z-scored. Functional connectivity was then computed as the Pearson correlation between all pairs of regional time series, yielding a $100\times100$ undirected connectivity matrix per participant. Each participant was summarized by the upper triangle of this matrix, which produced $100\times99/2=4{,}950$ features per participant. All downstream models used this 100-region, 4,950-feature representation.
\subsection{Classification targets}

Thirteen binary comparisons were evaluated in FOR2107 (Table~\ref{tab:app_targets}). BDI-I and SCL-90-R Global Severity Index (GSI) scores were divided at the median separately within HC and MDD. Ties were assigned to the low-symptom group. The resulting groups are sample-relative strata and not clinical cutoffs. Missing symptom measurements were excluded only from comparisons requiring that instrument.

\begin{table}[ht]
\centering
\caption{Empirical classification targets and available samples.}
\label{tab:app_targets}
\begin{tabular}{p{0.25\textwidth}p{0.42\textwidth}p{0.25\textwidth}}
\toprule
Domain & Definition & Available sample or class counts\\
\midrule
Recorded sex & Female versus male & 1,232 versus 720\\
Age 30 & $<30$ versus $\geq30$ years & 887 versus 1,065\\
Age 40 & $<40$ versus $\geq40$ years & 1,233 versus 719\\
Age 50 & $<50$ versus $\geq50$ years & 1,506 versus 446\\
Diagnosis & HC versus MDD & 1,129 versus 823\\
BDI-I strata & Four HC/MDD contrasts formed from diagnosis-specific low/high median splits & 60 missing; HC median 3; MDD median 15\\
SCL-90-R GSI strata & Four HC/MDD contrasts formed from diagnosis-specific low/high median splits & 57 missing; HC median 0.167; MDD median 0.811\\
\bottomrule
\end{tabular}
\end{table}

Fourteen binary comparisons were evaluated in SRPBS (Table~\ref{tab:app_targets_srpbs}). These comprised recorded sex; age below versus above 30, 40, and 50 years; all six pairwise contrasts among HC, MDD, ASD, and SSD; and four BDI contrasts combining diagnosis-specific low/high strata in HC and MDD. BDI was available for 418 HC and 190 participants with MDD. Diagnosis-specific medians were 5 in HC and 27.5 in MDD, with ties assigned to the low-symptom group. As in FOR2107, these are descriptive sample-relative strata rather than clinical cutoffs.

\begin{table}[ht]
\centering
\caption{SRPBS classification targets and available samples.}
\label{tab:app_targets_srpbs}
\begin{tabular}{p{0.25\textwidth}p{0.40\textwidth}p{0.22\textwidth}}
\toprule
Domain & Definition & Class counts\\
\midrule
Recorded sex & Female versus male & 394 versus 506\\
Age 30 & $<30$ versus $\geq30$ years & 239 versus 661\\
Age 40 & $<40$ versus $\geq40$ years & 473 versus 427\\
Age 50 & $<50$ versus $\geq50$ years & 697 versus 203\\
Pairwise diagnosis & HC--MDD; HC--ASD; HC--SSD; MDD--ASD; MDD--SSD; ASD--SSD & 500 HC; 220 MDD; 125 ASD; 55 SSD\\
BDI strata & MDD high--HC low; MDD high--HC high; MDD low--HC low; MDD low--HC high & HC low/high: 216/202; MDD low/high: 95/95\\
\bottomrule
\end{tabular}
\end{table}

\section{Empirical maximum-entropy and machine-learning analysis}
\label{app:maxent_methods}

This section details the fold-local generative and discriminative pipelines compared in Fig.~\ref{fig:main_results}A,B; their additional empirical results are reported in Appendix~\ref{app:supplementary_results}.

\subsection{End-to-end validation design}

Each of the 13 FOR2107 outcomes and 14 SRPBS outcomes was evaluated using five independently seeded repeats of stratified five-fold cross-validation, yielding 25 held-out estimates per outcome. Every data-dependent operation was restricted to the outer training fold: feature scaling, autoencoder fitting, latent-state construction, topology learning, class-conditional maximum-entropy estimation, preprocessing for the discriminative classifier, and classifier fitting. Held-out participants were used only for evaluation. This fold-local design prevents leakage from representation learning and graph selection.

Fold AUCs were averaged across the 25 held-out evaluations. Ninety-five percent confidence intervals were calculated from 10,000 class-stratified subject-level bootstrap samples while retaining the repeated held-out predictions belonging to each resampled participant. A maximum-entropy fit was considered numerically valid only when the optimizer reported convergence, all held-out scores were finite, and the maximum absolute gradient did not exceed $10^{-3}$. Optimization was initially limited to 500 iterations. In a post hoc numerical sensitivity check, each fold that did not meet these criteria was refitted once with a limit of 2,000 iterations, using the same data split, learned representation, model settings, and random seed. Retry was determined solely by numerical fit status and not by AUC. The retried results were adopted only if every refitted fold met the same criteria. Final summaries used the original predictions for initially valid folds and the replacement predictions for retried folds; confidence intervals were recalculated from these final predictions.

\subsection{Binary autoencoder}

Within each outer training fold, the 4,950 connectivity features were centered and scaled using training-participant statistics. A symmetric autoencoder mapped the standardized vector through a 240-unit nonlinear hidden layer to 120 bounded latent activations. The decoder mirrored the encoder. During the forward pass, each activation was thresholded at zero to form a hard binary state; a straight-through gradient estimator enabled optimization through the discrete bottleneck. Reconstruction was computed from the binary state itself, ensuring that the representation fitted by maximum entropy was also the information available to reconstruct connectivity.

The autoencoder minimized mean squared reconstruction error for 100 epochs using mini-batches of 128. Negative activations and non-negative activations defined the two binary values. The fold-specific continuous activations, thresholded states, and decoded connectivity were retained for the comparison analyses.

\subsection{Sparse tree-structured pairwise maximum entropy}

For $d=120$ binary variables $\boldsymbol z\in\{0,1\}^d$ and a selected edge set $\mathcal E$, the class-conditional model was
\begin{align}
p_g(\boldsymbol z)
&=\frac{\exp[-E_g(\boldsymbol z)]}{Z_g},\\
E_g(\boldsymbol z)
&=-\sum_{i=1}^{d}h_i^{(g)}z_i
-\sum_{(i,j)\in\mathcal E}J_{ij}^{(g)}z_i z_j.
\label{eq:app_empirical_maxent}
\end{align}
Within each training fold, states from both classes were pooled without labels to estimate smoothed pairwise mutual information. A maximum-information spanning tree was then learned. The resulting treewidth-one graph connected all 120 nodes with 119 edges. The same topology was imposed on both classes so that their comparison reflected class-conditional parameter differences rather than different graph complexity.

Each class model contained 120 node biases and 119 edge interactions, or 239 parameters per class and 478 parameters for the two-class comparison. Parameters were estimated by weakly regularized maximum likelihood. Because the topology was a tree, junction-tree or sum-product message passing yielded exact partition functions, node and edge marginals, likelihoods, and independent samples. Equivalently, a strictly positive tree distribution can be factorized from its consistent marginals as
\begin{equation}
p_g(\boldsymbol z)
=\prod_i p_g(z_i)
\prod_{(i,j)\in\mathcal E}
\frac{p_g(z_i,z_j)}{p_g(z_i)p_g(z_j)}.
\label{eq:app_tree_factorization}
\end{equation}
No unnormalized energy score or approximate Monte Carlo partition function was used for the empirical likelihood ratio.

\subsection{Held-out score and AUC}

For a held-out state $\boldsymbol z$, the fitted log likelihood ratio was
\begin{equation}
s(\boldsymbol z)
=\log\frac{\widehat p_1(\boldsymbol z)}{\widehat p_0(\boldsymbol z)}
=E_0(\boldsymbol z)-E_1(\boldsymbol z)
+\log\frac{Z_0}{Z_1}.
\label{eq:app_empirical_score}
\end{equation}
The partition-function term affects calibrated probabilities but is constant across participants and therefore does not affect the ROC curve. Held-out AUC was computed using Eq.~\eqref{eq:app_auc_pairwise}. Under the two fitted distributions, this is the optimal ranking score; on empirical data it remains conditional on the learned binary representation and the tree-pairwise approximation.

\subsection{Discriminative comparison pipelines}

A radial-basis-function support-vector classifier (RBF-SVC) was evaluated on four fold-specific representations:
\begin{enumerate}
\item the original 4,950 connectivity features;
\item the 120 continuous autoencoder activations;
\item the 120 thresholded binary latent states; and
\item the 4,950 connectivity features decoded from the binary states.
\end{enumerate}
Mean imputation, robust scaling, and classifier fitting were confined to the outer training fold. The SVC regularization parameter was $C=1$, and the kernel scale was the reciprocal of the number of input features. The binary-latent pipeline is the closest discriminative comparison with the maximum-entropy score; original connectivity can retain information absent from the compressed representation.

Three summaries compared each classifier profile with the maximum-entropy profile. The mean AUC difference was
\begin{equation}
\Delta\AUC=\frac{1}{K}\sum_{k=1}^{K}
(\AUC_{\mathrm{ML},k}-\AUC_{\mathrm{ME},k}),
\qquad K=13\text{ for FOR2107 and }K=14\text{ for SRPBS}.
\end{equation}
Pearson and Spearman correlations summarized agreement across comparison-level AUCs. Because comparisons reused participants and outcomes, these correlations were treated as descriptive rather than as evidence from independent replications. Relative reference attainment was defined above chance as
\begin{equation}
\eta_k=\frac{\AUC_{\mathrm{ML},k}-0.5}
{\AUC_{\mathrm{ME},k}-0.5}.
\label{eq:app_attainment}
\end{equation}
Here $\eta=1$ means equal discrimination above chance; values above one are possible because the empirical maximum-entropy result is a conditional reference rather than a universal bound.

\subsection{Exact 22-variable full-pairwise comparator}

A separate end-to-end pipeline compressed connectivity to 22 binary latent variables. Within each class, it fitted all 22 node biases and all $\binom{22}{2}=231$ pairwise interactions, for 253 parameters per class. All $2^{22}=4{,}194{,}304$ states were enumerated, so the partition function and the full discrete landscape were exact. This comparison changes both latent dimensionality and graph structure: it contrasts complete pipelines and does not isolate the effect of sparse versus dense interactions.

The two pipelines trade representation size against dependence structure. Full enumeration requires evaluating $2^d$ binary configurations and is practical for the 22-variable model, but becomes impractical at 120 variables. Restricting pairwise interactions to a tree allowed us to retain the larger representation while evaluating its partition function and likelihoods exactly by message passing. The approximation therefore concerns which dependencies the model can represent, rather than numerical approximation of its normalization. The 22-variable model retains every pairwise interaction within a more compressed representation; the 120-variable model retains more latent variables while restricting their dependence structure.

\subsection{Model adequacy and representation diagnostics}
\label{app:marginal_diagnostics}

Maximum-entropy adequacy was assessed using held-out agreement between predicted and observed first-order and pairwise marginals, the total latent-activity distribution, held-out likelihood, Brier score, and classification log loss. For each comparison, outer fold, and class, empirical marginals were calculated from held-out participants encoded by the training-fitted autoencoder. For binary variables, the first-order marginal $\E[Z_i]=\Pr(Z_i=1)$ is the activation probability of one latent unit, and the pairwise marginal $\E[Z_iZ_j]=\Pr(Z_i=1,Z_j=1)$ is the joint activation probability of two units. These are probabilities rather than correlations between latent units. The diagnostics included all 120 first-order marginals and all $\binom{120}{2}=7{,}140$ pairwise marginals, including pairs without a direct edge in the fitted tree.

Corresponding model marginals were estimated from 5,000 independent samples per class and fold drawn from the training-fitted tree distribution. Thus, the displayed predictions are Monte Carlo estimates of model expectations; the normalized likelihoods used for classification were evaluated exactly. Agreement was summarized separately for each marginal type by Pearson correlation and root-mean-square error across marginals within each class and fold, followed by averaging across classes, folds, repeats, and comparisons. The reported fractions within two sampling standard errors used the held-out empirical binomial standard error, with a minimum of $0.5/n$, where $n$ is the held-out class size. These fractions are descriptive diagnostics and do not include uncertainty from model training or from the sampled model expectations.

The marginal figures pool both classes and all 25 held-out folds for each comparison. Every first-order marginal is shown; for each class and fold, 120 pairwise marginals were selected without replacement using a fixed random seed for display. Numerical summaries use all 7,140 pairs. Each point therefore represents a marginal estimate, not an individual participant. These checks assess reproduction of held-out activation and co-activation probabilities; they do not establish agreement of the full joint distribution or calibration of individual diagnostic probabilities.

Autoencoder fidelity was summarized by participant-level Pearson correlation, root mean square error, and coefficient of determination between observed and decoded connectivity. In-sample full-cohort reconstruction summaries were treated as descriptive; every classification estimate derived from fold-specific models.

\section{Fixed-source synthetic benchmark}
\label{app:synthetic_methods}

\subsection{Purpose and source distributions}

The synthetic analysis tested the ceiling interpretation under correct model specification and provides the methods for Fig.~\ref{fig:main_results}C. Completed empirical autoencoders, connectivity scaling parameters, class-conditional maximum-entropy parameters, and graph topologies were selected as fixed sources and were not relearned when constructing the source distributions. Two source families were examined: the 120-variable tree with 119 edges and the 22-variable full-pairwise model with 231 edges. Their historical held-out diagnosis AUCs before synthetic generation were 0.628 and 0.582, respectively. These values describe the selected source fits, not the finalized repeated-cross-validation estimates in Table~\ref{tab:app_main_auc}. Samples generated from these fitted sources need not reproduce held-out human AUCs.

\subsection{Controlled landscape deformation}

For each source family, the HC natural parameters $\boldsymbol\theta_0$ and MDD natural parameters $\boldsymbol\theta_1$ defined the path in Eq.~\eqref{eq:app_theta_path}. Thus $\alpha=0$ reproduced the HC source, $\alpha=1$ reproduced the fitted MDD source, $0<\alpha<1$ interpolated between them, and $\alpha>1$ extrapolated along the same direction. The grid used increments of 0.1 from 0 to 2 and increments of 0.5 from 2.5 to 7.

At every $\alpha$, five independent datasets were generated, each containing 1,000 reference-class and 1,000 target-class observations. The tree model supplied exact independent junction-tree samples; the 22-variable model was sampled from its exactly enumerated categorical distribution. Binary samples were additionally decoded to 4,950 connectivity features when a reconstructed-connectivity analysis was required.

\subsection{Oracle and finite-sample AUC}

Because the source distributions were known, the source log likelihood ratio defined the population Bayes score. For the 22-variable model, its AUC was evaluated exactly over the enumerated state space:
\begin{equation}
\AUC_{\mathrm{oracle}}
=\sum_{\boldsymbol z_1,\boldsymbol z_0}
p_{\alpha}(\boldsymbol z_1)p_0(\boldsymbol z_0)
\left[
\mathbb I\{s(\boldsymbol z_1)>s(\boldsymbol z_0)\}
+\frac{1}{2}\mathbb I\{s(\boldsymbol z_1)=s(\boldsymbol z_0)\}
\right].
\label{eq:app_exact_discrete_auc}
\end{equation}
For the 120-variable tree, population AUC was estimated using ten independent Monte Carlo evaluations with 20,000 observations per class. A finite-sample oracle was also computed within each 2,000-observation replicate by applying the known source likelihood ratio to the sampled observations. This is an estimate of the population optimum, not a strict upper bound on every realized sample AUC.

\subsection{Round-trip refitting and machine-learning evaluation}

In a separate round-trip analysis, synthetic latent states were decoded into functional-connectivity space, passed back through the fixed encoder and binary bottleneck, and used to refit class-conditional maximum-entropy models. Held-out likelihood-ratio AUC from these refitted models quantified the loss caused by the decode--encode--refit process relative to the finite-sample source oracle.

The RBF-SVC pipeline displayed in Fig.~\ref{fig:main_results}C was evaluated using five-fold cross-validation on native binary states. The pipeline was also evaluated on decoded connectivity. These analyses used the same preprocessing definitions as the empirical comparison. Uncertainty intervals were based on Student's $t$ distribution across the five independent synthetic datasets; folds within a dataset were not treated as independent replicates.

\section{Supplementary empirical and synthetic results}
\label{app:supplementary_results}

The following results extend the FOR2107 empirical comparison in Fig.~\ref{fig:main_results}A, the SRPBS multi-disorder comparison in Fig.~\ref{fig:main_results}B, and the synthetic benchmark in Fig.~\ref{fig:main_results}C.

\subsection{Representation fidelity}

In the descriptive full-cohort analysis, decoding continuous 120-dimensional activations gave a mean participant-level correlation of 0.973 between observed and reconstructed connectivity, RMSE of 0.043, and $R^2$ of 0.944. After binary quantization, the corresponding values were 0.958, 0.054, and 0.912. All 120 binary units were active, the mean activation probability was approximately 0.50, and every participant had a unique binary code. These values describe representation fidelity and are not substituted for fold-local predictive estimates.

\subsection{Held-out maximum-entropy estimates and adequacy}
\label{app:heldout_adequacy}

The maximum-entropy reference was highest for age, followed by recorded sex, and substantially lower for diagnosis and symptom strata. Exact values available for the principal outcomes are listed in Table~\ref{tab:app_main_auc}. Family-level comparison with the 22-variable pipeline is shown in Table~\ref{tab:app_family_auc}.

\begin{table}[ht]
\centering
\caption{Principal held-out AUC estimates from the 120-variable tree maximum-entropy model.}
\label{tab:app_main_auc}
\begin{tabular}{lcc}
\toprule
Outcome & AUC & 95\% CI\\
\midrule
Age $<30$ versus $\geq30$ & 0.770 & 0.752--0.787\\
Age $<40$ versus $\geq40$ & 0.804 & 0.787--0.821\\
Age $<50$ versus $\geq50$ & 0.796 & 0.777--0.814\\
Female versus male & 0.714 & 0.694--0.733\\
HC versus MDD & 0.613 & 0.593--0.634\\
Eight symptom-defined contrasts & 0.598--0.628 & ---\\
\bottomrule
\end{tabular}
\end{table}

\begin{table}[ht]
\centering
\caption{Mean held-out maximum-entropy AUC by outcome family.}
\label{tab:app_family_auc}
\begin{tabular}{lccc}
\toprule
Outcome family & Comparisons & 120-variable tree & 22-variable full pairwise\\
\midrule
Demographic and diagnostic & 5 & 0.739 & 0.675\\
BDI-I strata & 4 & 0.614 & 0.576\\
SCL-90-R GSI strata & 4 & 0.609 & 0.566\\
All outcomes & 13 & 0.661 & 0.611\\
\bottomrule
\end{tabular}
\end{table}

All 325 fold-level two-class sparse comparisons passed the numerical validity criteria. Across held-out folds and classes, predicted and observed first-order marginals correlated at a mean $r=0.822$, and pairwise marginals at $r=0.738$. In total, 91.3\% of first-order marginals and 86.9\% of pairwise marginals fell within two sampling standard errors of their observations. The mean Jensen--Shannon divergence for the total-activity distribution was 0.059. Probability calibration was weaker than ranking performance: mean Brier score was 0.261 and mean classification log loss was 0.821. The primary evidence therefore concerns discrimination rather than calibrated individual risk probabilities.

\begin{figure}[p]
    \centering
    \includegraphics[width=\textwidth]{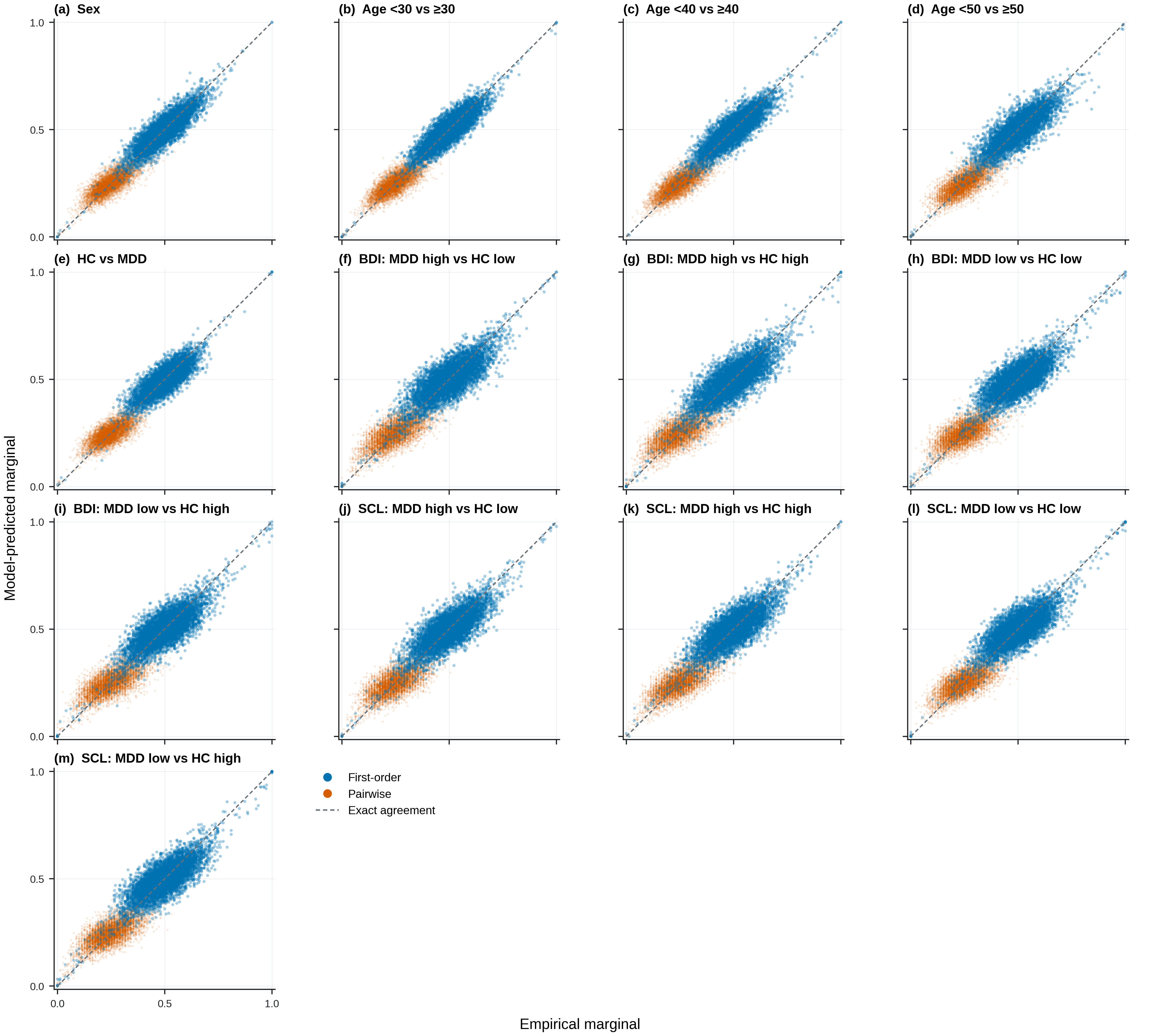}
    \caption{\textbf{Held-out marginal agreement for the pooled-cohort maximum-entropy analysis.}
    Panels (a--m) show the 13 comparisons from Fig.~\ref{fig:main_results}A. The horizontal axis gives empirical marginals from held-out participants; the vertical axis gives corresponding estimates from 5,000 independent samples per class and fold from the training-fitted 120-variable tree model. Blue points represent first-order activation probabilities and orange points pairwise joint activation probabilities. Each panel pools both classes and five repeats of five-fold cross-validation. All first-order marginals are displayed; 120 of the 7,140 pairwise marginals per class and fold are shown using fixed-seed sampling. Numerical summaries retain all pairs, including pairs without direct tree edges. The dashed diagonal denotes exact agreement. Methods and interpretation are provided in Appendices~\ref{app:marginal_diagnostics} and \ref{app:heldout_adequacy}.}
    \label{fig:pooled_marginal_agreement}
\end{figure}

Appendix Fig.~\ref{fig:pooled_marginal_agreement} displays the underlying held-out marginal agreement for each comparison, using the procedure described in Appendix~\ref{app:marginal_diagnostics}. The correspondence supports the fitted models' ability to reproduce first-order and pairwise structure beyond the training data, while the remaining dispersion indicates that this reproduction is approximate.

\subsection{Machine learning relative to the conditional reference}

In FOR2107, all four discriminative profiles closely tracked the maximum-entropy profile (Table~\ref{tab:app_ml_comparison}). The binary-state SVC was the closest direct comparison, with mean AUC 0.669 versus 0.661 for maximum entropy, Pearson $r=0.996$, and Spearman $\rho=0.967$. Decoded connectivity also closely matched the conditional reference. Original connectivity and continuous latent activations exceeded it on average, consistent with their retaining continuous information not represented by the thresholded binary tree model.

\begin{table}[ht]
\centering
\caption{RBF-SVC performance relative to the 120-variable maximum-entropy reference across 13 outcomes.}
\label{tab:app_ml_comparison}
\begin{tabular}{lrrrrr}
\toprule
Input representation & Features & Mean AUC & $\Delta$AUC & Pearson $r$ & Attainment\\
\midrule
Original connectivity & 4,950 & 0.712 & +0.051 & 0.971 & 129\%\\
Binary latent states & 120 & 0.669 & +0.008 & 0.996 & 104\%\\
Binary-decoded connectivity & 4,950 & 0.666 & +0.005 & 0.996 & 102\%\\
Continuous latent activations & 120 & 0.678 & +0.017 & 0.993 & 109\%\\
\bottomrule
\end{tabular}
\end{table}

Spearman correlations ranged from 0.967 to 0.984 across the four representations. The binary-latent classifier exceeded the fitted reference in 10 of 13 tasks, which is compatible with finite-sample variation and departures from the tree approximation. Original connectivity exceeded the reference in all 13 tasks, directly demonstrating why the empirical value cannot be interpreted as a bound on richer representations.

\subsection{Independent multi-disorder analysis in SRPBS}
\label{app:srpbs_results}

The SRPBS analysis repeated the complete comparison between the 120-variable tree maximum-entropy model and an RBF support-vector classifier applied to the same binary states. No fitted values were carried over from FOR2107. Feature scaling, autoencoder training, construction of the binary states, estimation of the tree structure and class distributions, and classifier training were all repeated within each cross-validation training set. The 14 comparisons and the number of participants in each group are listed in Table~\ref{tab:app_targets_srpbs}.

Across the 14 comparisons, the mean maximum-entropy AUC was 0.691 and the mean AUC of the RBF support-vector classifier was 0.706. The AUCs from the two methods showed close descriptive agreement (Pearson $r=0.975$; Spearman $\rho=0.974$), and the classifier AUC was higher by 0.015 on average. For the four demographic comparisons, the mean AUCs were 0.725 for maximum entropy and 0.739 for the classifier. The corresponding values were 0.717 and 0.736 for the six diagnostic comparisons, and 0.619 and 0.630 for the four BDI comparisons. Individual estimates, bootstrap intervals, and the numbers of model fits that met the numerical criteria are shown in Table~\ref{tab:app_srpbs_auc}.

\begin{table}[ht]
\centering
\scriptsize
\caption{Held-out SRPBS AUC estimates for the 120-variable tree maximum-entropy model and the RBF support-vector classifier applied to the same binary states. The 95\% bootstrap intervals were calculated by resampling participants separately within each class. The final column gives the number of maximum-entropy fits, out of 25 cross-validation folds, that met all predefined numerical criteria; these criteria do not apply to the support-vector classifier.}
\label{tab:app_srpbs_auc}
\setlength{\tabcolsep}{4pt}
\begin{tabular}{lccc}
\toprule
Comparison & Maximum-entropy AUC (95\% CI) & Classifier AUC (95\% CI) & Fits meeting criteria\\
\midrule
Recorded sex & 0.717 (0.688--0.745) & 0.736 (0.708--0.763) & 25/25\\
Age 30 & 0.717 (0.687--0.747) & 0.729 (0.701--0.756) & 25/25\\
Age 40 & 0.740 (0.712--0.767) & 0.764 (0.737--0.789) & 25/25\\
Age 50 & 0.726 (0.691--0.759) & 0.725 (0.693--0.757) & 25/25\\
HC--MDD & 0.642 (0.605--0.677) & 0.639 (0.606--0.672) & 25/25\\
HC--ASD & 0.777 (0.744--0.808) & 0.797 (0.766--0.828) & 25/25\\
HC--SSD & 0.741 (0.689--0.792) & 0.737 (0.679--0.794) & 25/25\\
MDD--ASD & 0.827 (0.791--0.862) & 0.856 (0.820--0.889) & 25/25\\
MDD--SSD & 0.705 (0.653--0.756) & 0.713 (0.657--0.769) & 25/25\\
ASD--SSD & 0.611 (0.539--0.681) & 0.675 (0.600--0.749) & 25/25\\
MDD high--HC low BDI & 0.589 (0.538--0.639) & 0.591 (0.540--0.641) & 25/25\\
MDD high--HC high BDI & 0.539 (0.496--0.583) & 0.556 (0.506--0.606) & 25/25\\
MDD low--HC low BDI & 0.676 (0.625--0.726) & 0.689 (0.639--0.739) & 25/25\\
MDD low--HC high BDI & 0.670 (0.613--0.724) & 0.682 (0.627--0.737) & 25/25\\
\bottomrule
\end{tabular}
\end{table}

In the initial 500-iteration analysis, 241 of 350 folds met every numerical criterion. The remaining 109 were flagged only because at least one class-specific optimizer reached the iteration limit; all produced finite held-out scores, and all maximum absolute gradients remained below $10^{-3}$. Each flagged fold was refitted once with a 2,000-iteration limit, and all 109 then met the criteria. The final summaries therefore use the original predictions for 241 folds and the replacement predictions for 109 folds. Relative to the initial all-finite summary, the mean absolute change in comparison-level AUC was 0.00021 and the maximum was 0.00064; no confidence-interval endpoint changed by more than 0.00091. Pearson $r$ changed from 0.9752 to 0.9749, while Spearman $\rho$ remained 0.9736.

The multisite design also changes what diagnostic separability means. HC participants were represented across eight study sites and participants with MDD across five, whereas ASD and SSD participants occurred only at the SWA and UTO sites. Consequently, scanner, protocol, recruitment, and diagnosis could not be distinguished fully in these data. The high AUCs for some comparisons between diagnostic groups may therefore reflect site and sample composition as well as disorder-related information. The clear finding is that maximum-entropy and RBF-SVC estimates agree when both methods are applied to the same binary brain-state data within each cross-validation fold. Because diagnostic groups were distributed unevenly across sites, these AUCs should not be assumed to generalize to other sites or samples.

\subsection{High-dimensional sparse versus low-dimensional full-pairwise models}

Both pipelines began with the original 4,950-element functional-connectivity vector. The 120-variable tree model exceeded the 22-variable full-pairwise model in all 13 comparisons, with a mean AUC advantage of 0.050 and a range of 0.023--0.083. Separate descriptive full-cohort reconstructions also showed greater fidelity after binary encoding at 120 variables than at 22 variables, with mean participant-level correlations of 0.958 and 0.877, respectively. These reconstruction values describe the full-cohort networks, rather than held-out reconstruction by the fold-specific autoencoders. Because latent dimensionality, autoencoder capacity, and interaction structure differed, the comparison does not isolate their individual contributions. The results are consistent with the larger representation preserving useful connectivity information despite the restriction to a tree.

The discriminative analyses showed the same advantage for information passed through the binary bottleneck. Mean RBF-SVC AUC was 0.669 versus 0.632 on the 120- and 22-variable binary representations, and 0.666 versus 0.624 on connectivity decoded from those representations. Both comparisons favored the 120-variable pipeline in every outcome. The classifier applied directly to original connectivity provided a common baseline, with mean AUC 0.712 in both analyses, because this pipeline bypassed the autoencoder. Its higher performance shows that even the 120-variable representation omitted information useful for classification. The advantage over 22 variables therefore concerns the information retained by the compressed pipelines.

\subsection{Choice of latent dimensionality and treewidth}
\label{app:latent_choice}

We retained $d=120,w=1$ to give priority to reproduction of the latent distributions while preserving useful discrimination. A separate configuration analysis compared sparse models with 20--480 latent variables and treewidths 1--5 across diagnosis, recorded sex, and the age-40 comparison. Scaling, autoencoder training, topology learning, and parameter estimation were confined to the corresponding training splits. Fit was assessed through held-out first-order activation probabilities and co-activation probabilities on the selected graph edges. Residuals were standardized by their sampling standard errors, and root-mean-square errors were averaged across moment types, classes, folds, and outcomes. This criterion measures marginal agreement rather than agreement of the complete joint distribution.

The fit-priority rule retained configurations within one standard error of the best marginal fit, then ranked them by equally weighted AUC, with complexity used to break ties. It selected $d=120,w=1$. The rules giving greater priority to AUC selected $d=240,w=1$. In this configuration analysis, the mean standardized marginal error was 1.158 at 120 variables and 1.206 at 240 variables. The larger representation therefore offered somewhat better discrimination at the cost of poorer fit under this criterion. These configuration comparisons support the rationale for the reported model; their results are distinct from the repeated cross-validation estimates of the fixed pipelines.

The fixed-pipeline analysis across all 13 outcomes showed modest gains from doubling the latent size: mean maximum-entropy AUC increased from 0.661 to 0.672, binary-latent SVC AUC from 0.669 to 0.680, and decoded-connectivity SVC AUC from 0.666 to 0.675. Gains were larger for age and recorded sex; HC versus MDD increased from 0.613 to 0.619. Increasing treewidth from 1 to 2 at 120 variables slightly improved pairwise marginal reproduction but did not improve discrimination: mean maximum-entropy AUC decreased from 0.661 to 0.657. We therefore retained the tree as the simpler dependence model, and 120 variables as a compromise between marginal fit and the additional discrimination available at larger latent sizes.

\subsection{Synthetic benchmark}

At $\alpha=0$, all methods were near chance. Oracle AUC increased monotonically with separation and approached one for both source families. The finite-sample oracle closely followed the population oracle, indicating that 1,000 observations per class were sufficient to estimate source-defined separability over the tested grid. Selected values are shown in Table~\ref{tab:app_synthetic_results}.

\begin{table}[ht]
\centering
\caption{Selected synthetic AUC results.}
\label{tab:app_synthetic_results}
\begin{tabular}{lccccc}
\toprule
Source & $\alpha$ & Population oracle & Finite oracle & ML latent & ML decoded\\
\midrule
120-variable tree & 0.5 & 0.654 & 0.649 & 0.596 & 0.594\\
120-variable tree & 1.0 & 0.786 & 0.781 & 0.720 & 0.703\\
120-variable tree & 1.5 & 0.885 & 0.884 & 0.828 & 0.801\\
120-variable tree & 2.0 & 0.945 & 0.943 & 0.903 & 0.873\\
22-variable full pairwise & 0.5 & 0.627 & 0.629 & 0.557 & 0.552\\
22-variable full pairwise & 1.0 & 0.741 & 0.732 & 0.634 & 0.629\\
22-variable full pairwise & 1.5 & 0.835 & 0.831 & 0.740 & 0.729\\
22-variable full pairwise & 2.0 & 0.905 & 0.905 & 0.825 & 0.808\\
\bottomrule
\end{tabular}
\end{table}

At the MDD endpoint ($\alpha=1$), the 120-variable population oracle was 0.786 and its finite-sample estimate was 0.781; latent-state and decoded-connectivity SVCs reached 0.720 and 0.703. For the 22-variable source, the corresponding values were 0.741, 0.732, 0.634, and 0.629. Between $\alpha=0.1$ and 2.0, latent-state ML averaged 0.055 AUC below the 120-variable oracle and 0.082 below the 22-variable oracle; decoded-connectivity deficits were 0.073 and 0.088. These results test classification against a known population optimum; they do not estimate a human metabolic or diagnostic ceiling.

\subsection{Replication in non-overlapping acquisition-center subsets}
\label{app:site_replication}

To assess whether the pooled findings depended on combining acquisition centers, we repeated the empirical and fixed-source synthetic analyses separately in Marburg and M\"unster. The two subsets contained 1,006 participants in Marburg (642 HC and 364 MDD) and 946 in M\"unster (487 HC and 459 MDD). They comprised different participants acquired by separate local teams using the scanners described in Appendix~\ref{app:empirical_methods}. All empirical preprocessing of the connectivity features, autoencoder fitting, binary-state construction, topology learning, and class-conditional estimation were repeated within the outer training folds of the respective center. Each center used the same 13 comparison definitions and five repeats of stratified five-fold cross-validation. Diagnosis-specific symptom medians were recomputed within each center, so the high/low strata were relative to that center's sample. These analyses evaluate reproducibility within two disjoint subsets of the pooled cohort; they do not evaluate transfer of a model trained at one center to participants at the other.

\begin{figure}[p]
    \centering
    \includegraphics[width=\textwidth,height=0.76\textheight,keepaspectratio]{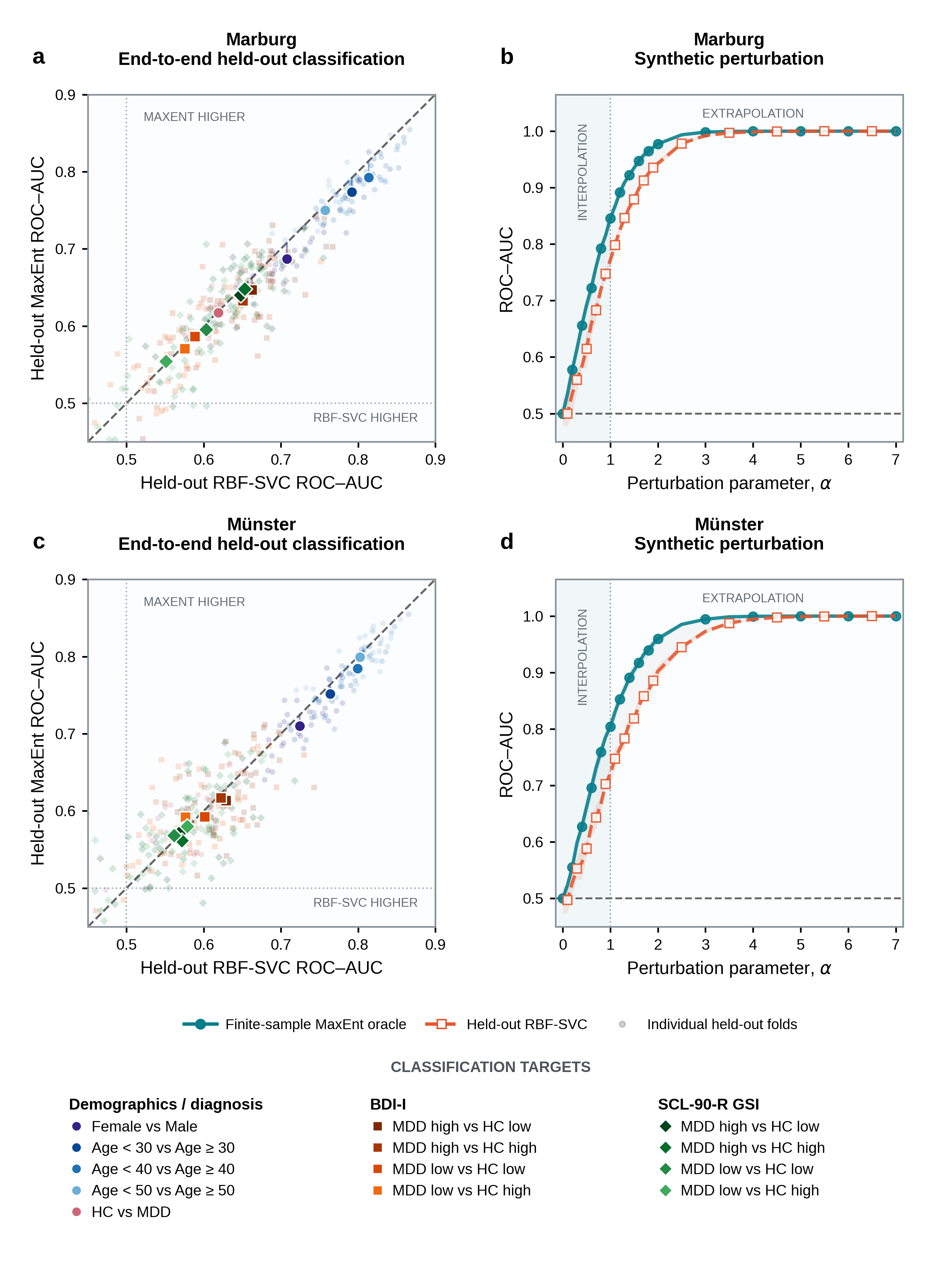}
    \caption{\textbf{City-specific replication of the empirical and synthetic comparisons in Fig.~\ref{fig:main_results}.}
    \textbf{a,c,} Held-out maximum-entropy AUC versus the AUC of an RBF support-vector classifier applied to the same binary states in Marburg and M\"unster. Large points show means for 13 comparisons; translucent points show results from 25 cross-validation folds per comparison. The diagonal denotes equality. For each center, every stage of the analysis was trained separately within each cross-validation training set.
    \textbf{b,d,} Simulations based on HC and MDD distributions fitted separately in Marburg and M\"unster. Curves show the mean AUC of the best ranking calculated from each simulated sample and the held-out RBF-SVC AUC; colored bands show 95\% Student's $t$ intervals across five generated datasets per $\alpha$. Each dataset contains 1,000 observations per class. The two centers reproduce the broad empirical hierarchy and the increase in separability as the energy landscape is deformed. The empirical maximum-entropy AUCs apply only to the binary representation used here, whereas the known simulated distributions define the highest theoretically achievable AUCs. See Appendix~\ref{app:site_replication}.}
    \label{fig:site_replication}
\end{figure}

Both centers reproduced the broad hierarchy in Fig.~\ref{fig:main_results}A: age and recorded sex were more separable than diagnosis and symptom-defined contrasts. Across the 13 comparisons, the correlation between maximum-entropy and binary-latent RBF-SVC AUCs was $r=0.997$ in Marburg and $r=0.995$ in M\"unster (Appendix Fig.~\ref{fig:site_replication}a,c). Mean maximum-entropy and RBF-SVC AUCs were 0.653 and 0.663 in Marburg, and 0.641 and 0.645 in M\"unster. HC-versus-MDD maximum-entropy AUC was 0.617 and 0.588, respectively. Thus, the broad ordering and agreement between maximum-entropy and RBF-SVC AUCs were reproducible across centers, although individual AUCs and the ordering of closely performing contrasts varied.

\begin{figure}[p]
    \centering
    \includegraphics[width=\textwidth]{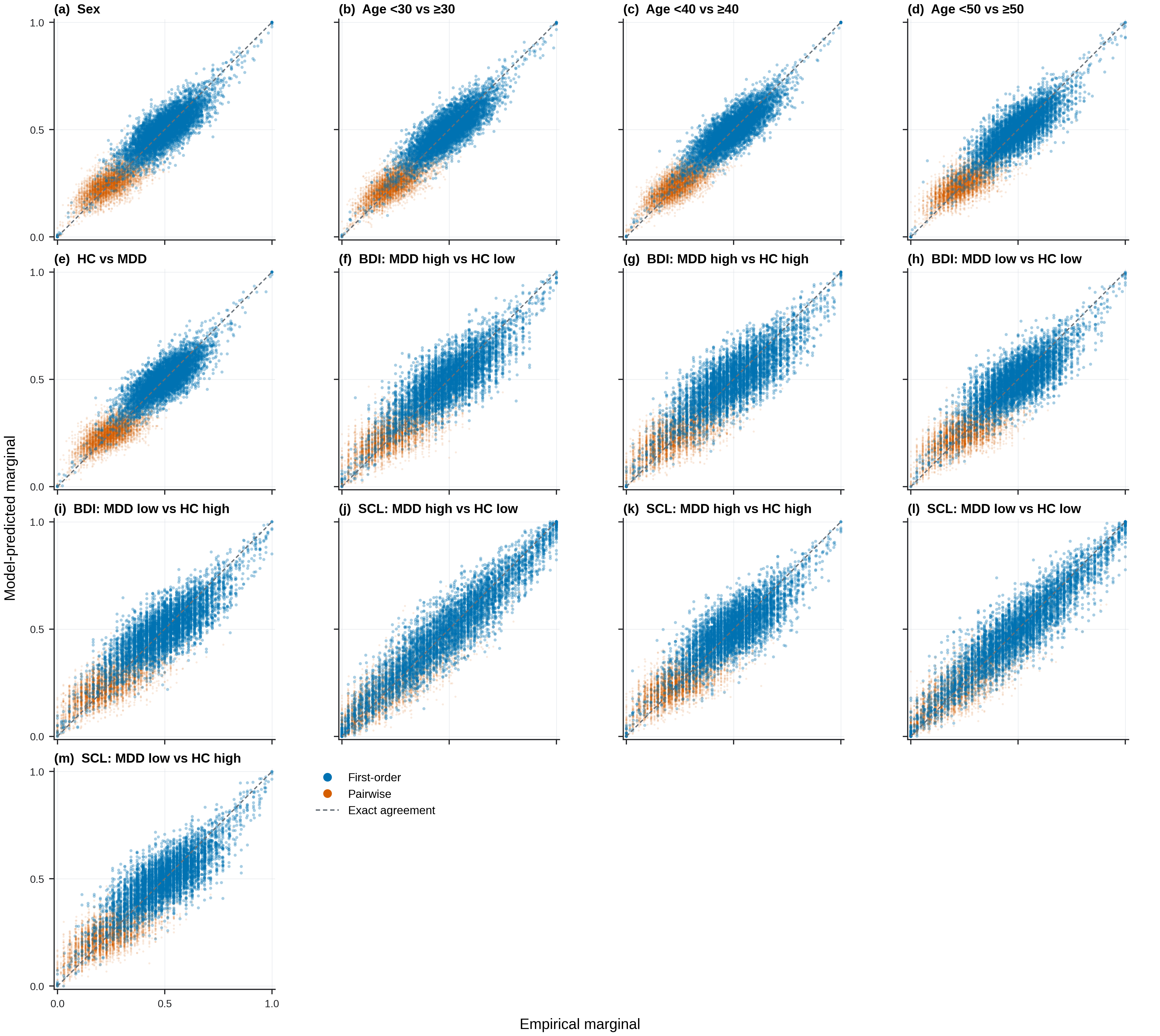}
    \caption{\textbf{Held-out marginal agreement in the Marburg subset.}
    Panels (a--m) show the 13 comparisons evaluated within Marburg ($n=1{,}006$ before comparison-specific exclusions). Blue and orange points compare held-out empirical first-order and pairwise marginals with estimates from the independently fitted Marburg models. Each panel pools both classes and 25 outer folds; graphical conventions, model sampling, and pairwise display sampling match Appendix Fig.~\ref{fig:pooled_marginal_agreement}. The diagonal denotes exact agreement. This analysis assesses whether marginal correspondence persists when representation learning and model fitting use only one acquisition center. Methods and numerical results are provided in Appendices~\ref{app:marginal_diagnostics} and \ref{app:site_replication}.}
    \label{fig:marburg_marginal_agreement}
\end{figure}

\begin{figure}[p]
    \centering
    \includegraphics[width=\textwidth]{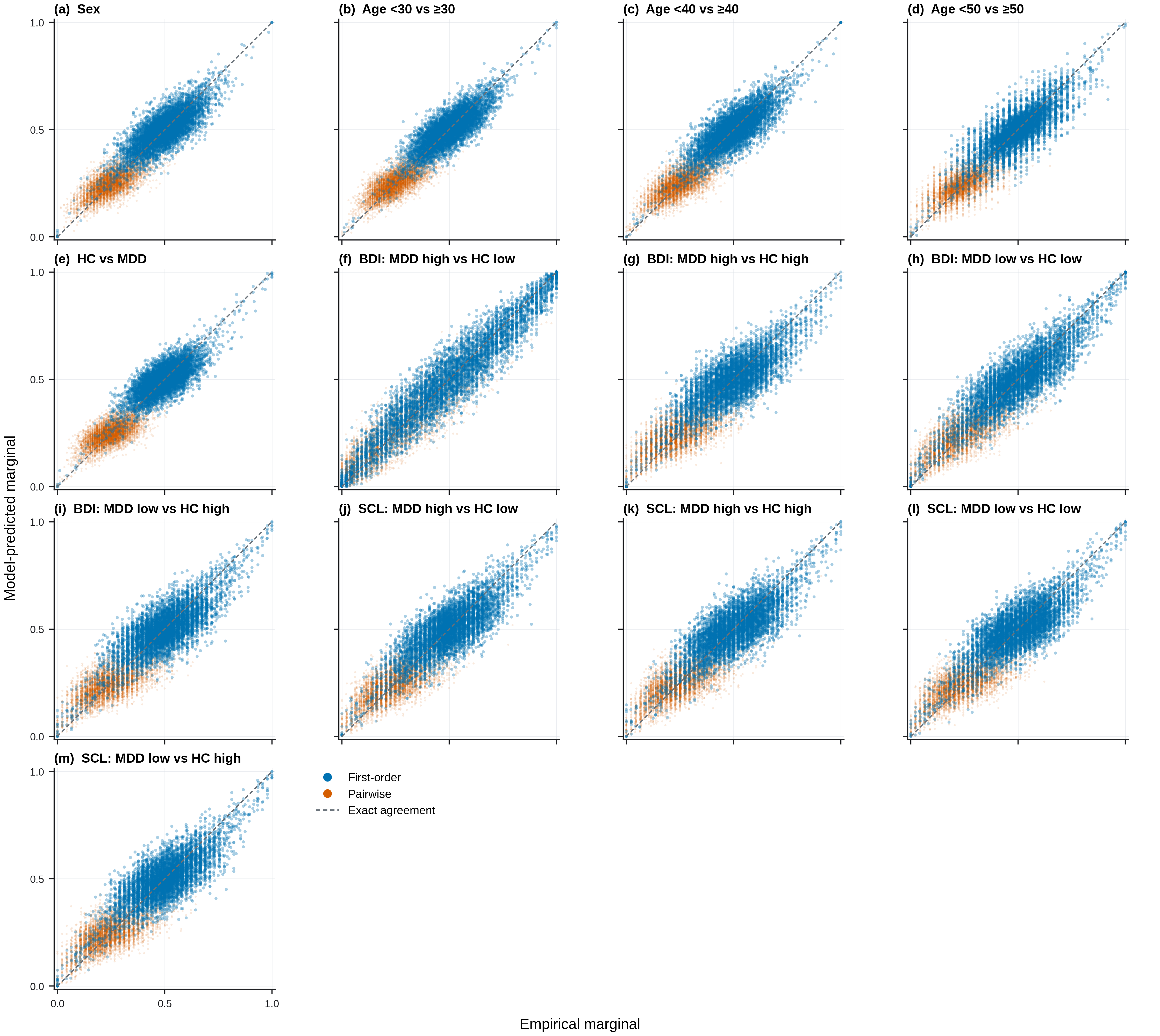}
    \caption{\textbf{Held-out marginal agreement in the M\"unster subset.}
    Panels (a--m) show the 13 comparisons evaluated within M\"unster ($n=946$ before comparison-specific exclusions). Blue and orange points compare held-out empirical first-order and pairwise marginals with estimates from the independently fitted M\"unster models. Each panel pools both classes and 25 outer folds; graphical conventions, model sampling, and pairwise display sampling match Appendix Fig.~\ref{fig:pooled_marginal_agreement}. The diagonal denotes exact agreement. No Marburg participants were used to learn these representations or fit these distributions. Methods and numerical results are provided in Appendices~\ref{app:marginal_diagnostics} and \ref{app:site_replication}.}
    \label{fig:munster_marginal_agreement}
\end{figure}

All 325 fold-level two-class maximum-entropy fits in each center passed the numerical validity criteria. Held-out marginal checks used the same procedure as the pooled analysis (Appendix~\ref{app:marginal_diagnostics}; Appendix Figs.~\ref{fig:marburg_marginal_agreement} and \ref{fig:munster_marginal_agreement}). Mean first-order and pairwise correlations were 0.843 and 0.775 in Marburg, and 0.818 and 0.747 in M\"unster. Corresponding mean root-mean-square errors were 0.072 and 0.068 in Marburg, and 0.073 and 0.069 in M\"unster, compared with 0.052 and 0.051 in the pooled cohort. These results show that substantial marginal correspondence persisted when each center was analyzed independently, while absolute errors were larger. Differences in dispersion can reflect smaller held-out class sizes, sample composition, and independently learned representations as well as model fit.

For synthetic replication, each center supplied its own full-subset autoencoder and HC/MDD tree distributions, fitted separately from those of the other center. These distributions defined center-specific natural-parameter paths using the same $\alpha$ grid, five independent datasets per grid point, and 1,000 observations per class described in Appendix~\ref{app:synthetic_methods}. The native binary-latent RBF-SVC was evaluated by five-fold cross-validation within each generated dataset. Both benchmarks reproduced the increase in separability with stronger landscape deformation and convergence toward an AUC of one (Appendix Fig.~\ref{fig:site_replication}b,d). At $\alpha=1$, finite-sample oracle and RBF-SVC AUCs were 0.846 and 0.772 for Marburg, and 0.804 and 0.721 for M\"unster. The mean classifier curves remained at or below the finite-sample oracle for positive $\alpha$, with the separation most evident at weak and intermediate deformations. These are benchmarks of separately fitted source distributions, rather than additional empirical estimates of diagnostic AUC.

\subsection{Interpretive boundary}

The mathematical and empirical analyses support two distinct conclusions. For data generated from the specified maximum-entropy distributions, the likelihood-ratio AUC is the highest theoretically achievable AUC. For FOR2107 and SRPBS, the fitted energy-difference AUC estimates the separability captured by one particular 120-variable tree model, with every step fitted using only the cross-validation training data. The agreement with RBF-SVC performance shows that the two methods give nearly the same ordering of comparison difficulty within each dataset. It does not show that metabolic expenditure has been measured, that SRPBS differences associated with study site are specific to diagnosis, or that the reported values limit every possible neuroimaging representation.

%

\section{Targeted evidence map of brain metabolism and diagnostic classification}
\label{app:evidence_map}

\subsection{Aim and status of the analysis}

This appendix documents the fourth step introduced in
Section~\ref{sec:introduction}, reported in
Section~\ref{sec:results_evidence}, and discussed in
Section~\ref{sec:discussion}. The purpose of this analysis was to
determine whether published aggregate evidence could test the
hypothesized association between the magnitude of a diagnosis-related
alteration in brain energy metabolism and the attainable performance of
a brain-based diagnostic classifier. We designed it as a targeted
evidence map with an exploratory quantitative component; it does not
attempt to review all metabolic abnormalities in psychiatry
systematically. The prespecified protocol was deposited on the Open
Science Framework \cite{kheirkhah2026targeted}
(\url{https://doi.org/10.17605/OSF.IO/PXWSK}).

The unit of analysis was a population-matched pair consisting of (i) a
standardized diagnosis-versus-healthy-control effect for a direct brain
energy/metabolism measure and (ii) a held-out
diagnosis-versus-healthy-control AUC based on brain-derived features. A
diagnosis-level correlation would have been considered only if
sufficiently comparable pairs met the predefined eligibility criteria
and the comparability criteria described below. Because no pair
satisfied all criteria, no metabolism--AUC meta-regression or
correlation was performed.
Supplementary Tables~S1--S8, including Tables~S1b and S8b, are provided
in the supplementary data deposit accompanying this article, not embedded in
this manuscript.

\subsection{Information sources and search strategy}

PubMed was the principal bibliographic source. The registered protocol
covered major depressive disorder, schizophrenia-spectrum and other
psychotic disorders, post-traumatic stress disorder, generalized anxiety
disorder, attention-deficit/hyperactivity disorder, and autism spectrum
disorder, and specified PubMed searches for recent metabolism and proxy
studies. The completed map additionally covers panic disorder, social
anxiety disorder, bipolar disorder, obsessive-compulsive disorder, and
clinical high risk for psychosis, and its coverage and supplementary
searches were run without a date limit; these extensions and the
quantitative-ASL coverage column are exploratory additions to the
registered scope. Conditions outside the
coverage grid were also included exploratorily, for two purposes:
selected additional psychiatric and neurological conditions illustrate
measurement dependence in Fig.~\ref{fig:metabolic_evidence_limits}, and
two of them, anorexia nervosa and Alzheimer's disease, also enter the
candidate-pair analysis.

The metabolism search combined diagnostic terms with terms for glucose
metabolism, cerebral glucose metabolism, FDG, FDG-PET,
fluorodeoxyglucose, positron-emission tomography, oxygen metabolism,
high-energy phosphates, ketone metabolism, and related measures. The
classifier search combined diagnosis and healthy-control terms with
brain-feature terms (MRI, functional MRI, functional connectivity, PET,
MRS, EEG, and fNIRS), classification terms, and ROC/AUC terms. Coverage
for Appendix Fig.~\ref{fig:evidence_map_coverage}A was mapped with a
grid of 1406 structured condition-by-measure queries (every
condition-term variant crossed with every measurement-term variant,
including each measurement class's plain modality name and standard
wordings of each condition), executed without a date limit; null results
were retained (901 of the 1406 queries returned no records). Every
record retrieved for a cell without eligible studies was screened at
title and abstract level, as was every meta-analysis retrieved for a cell
coded ``studies exist'', which could have supplied a usable pooled effect
(255 decisions on 228 records; Supplementary Table~S8b). Cell codes
were assigned from these screened records together with the extraction
records (Supplementary Table~S1) and studies located by targeted
supplementary searches (Supplementary Table~S1b, which records how each
study was found). Of the 87 independently located supporting records
that carry a PMID, the grid retrieved 75 for at least one of the cells
they support, and it retrieved at least one supporting record for 27 of
the 31 cells with eligible studies; the remaining four cells rest on
studies found by the supplementary searches. The query strings with per-query record counts are provided in
Supplementary Table~S8.

\subsection{Screening and eligibility}

Records were deduplicated and screened against the following criteria
at title and abstract level and then on the full text where it was
accessible, otherwise on the abstract; for the coverage grid, every
screening decision is deposited with its reason (Supplementary
Table~S8b). Human metabolism
studies were eligible if they compared a psychiatric group with healthy
controls using a brain-derived metabolic measure and supplied sufficient
quantitative or qualitative information to code the contrast. Classifier
studies were eligible if they used brain-derived features to distinguish
a diagnosis from healthy controls and reported an extractable ROC AUC or
AUROC. Studies focused only on treatment response, prognosis, symptom
severity, within-diagnosis subtypes, diagnosis-versus-diagnosis
classification, peripheral markers, or combined brain/peripheral models
without a separable brain-only estimate were excluded.

\subsection{Hierarchy of metabolic measures}

The preregistration defined a three-tier hierarchy of measures according
to their proximity to physical energy metabolism; the completed map
retained this hierarchy and extended Tier~1 to include ketone
metabolism:

\begin{enumerate}
  \item \textbf{Tier 1, direct metabolism or bioenergetics:} FDG-PET or
    absolute cerebral glucose metabolism; cerebral metabolic rate of
    oxygen (CMRO$_2$); ketone metabolism; ATP- and
    phosphocreatine-related measures; and $^{31}$P-MRS bioenergetics.
  \item \textbf{Tier 2, metabolic or neurochemical proxies:} glutamate,
    Glx, GABA, lactate, N-acetylaspartate, arterial-spin-labeling
    perfusion, and cerebral blood flow.
  \item \textbf{Tier 3, indirect energetic or landscape proxies:}
    TSPO-PET, neuroinflammation, glial or astrocytic markers,
    network-control energy, transition energy, and statistical
    energy-landscape measures.
\end{enumerate}

The attempted metabolism--AUC pairing was restricted to Tier 1.
Quantitative ASL cerebral blood flow was retained as a separate coverage
column because quantified flow that has not been normalized away can
support calibrated metabolic inference, but it was not treated as
interchangeable with glucose consumption, oxygen consumption, or
phosphate turnover. Lactate, a Tier-2 quantity, enters the coverage grid
only through the ketone/lactate column, and only when measured
dynamically (Appendix~\ref{app:evidence_coverage}).

\subsection{Extraction and harmonization rules}

For each metabolism record, the extraction captured diagnosis and
subgroup definition, healthy-control comparator, sample size, anatomical
scope, acquisition modality, physical quantity, absolute or normalized
units, effect direction, reported or calculable standardized effect, and
source identifier. Hedges' $g$ or Cohen's $d$ was retained when reported
or directly calculable from group means, standard deviations, and sample
sizes. Peak statistics, voxelwise or clusterwise test statistics, $t$ or
$F$ values without sufficient conversion information, and qualitative
statements were not treated as comparable diagnosis-level effect sizes.
For each classifier record, the extraction captured sample and site
structure, feature modality, validation design, classification target,
AUC definition, and held-out AUC.

Candidate pairs were formed by matching the two halves on diagnosis and,
where the literature allowed, on population definition and stage of
illness. A candidate pair was eligible for quantitative analysis only if
both halves referred to a compatible population, the metabolism
half was a standardized effect for a direct energetic measure expressed
on the scale shared by the other pairs (the same physical quantity, unit
convention, and spatial scope, so that diagnoses could be ordered), and
the classifier half reported the required held-out
diagnosis-versus-control AUC.

\subsection{Coverage analysis}
\label{app:evidence_coverage}

For Appendix Fig.~\ref{fig:evidence_map_coverage}A, each of 45 cells
(nine conditions by five measurement classes) was assigned one of three
codes: $2$, a usable pooled effect size in the column's units; $1$,
eligible studies existed but no comparable pooled scalar was available,
including where a pooled estimate existed but failed the column's
comparability criteria; or $0$, no eligible study was identified. The
five columns were FDG-PET glucose metabolism,
$^{31}$P-MRS ATP/phosphocreatine, quantitative ASL cerebral blood flow,
CMRO$_2$, and ketone/lactate dynamics. A study satisfied a cell only if it
measured the column's physical quantity in the brain relative to healthy
controls; a value derived from a different measurement or assessed
peripherally did not qualify. Values expressed relative to a reference
(a reference region, the whole brain, or the total phosphorus signal)
could establish that eligible studies existed, but a pooled effect in
such units could not by itself make a cell code~2. For the ASL column, a cell
required at least one primary study that quantified resting cerebral
blood flow with ASL in patients and healthy controls without restricting
the comparison to globally normalized or Z-scored flow; meta-analyses,
reviews, and non-ASL flow measurements did not by themselves qualify a
cell. The ketone/lactate column required brain ketone or lactate to be
measured dynamically, that is, compared between two states (for example
during a task or metabolic challenge and at rest) or measured by isotopic
turnover; a single-state concentration did not qualify.

Only one of the 45 cells, schizophrenia FDG-PET, contained a usable
pooled metabolic effect. Eight of the nine conditions had some FDG-PET
evidence, eight had $^{31}$P-MRS evidence, eight had quantitative-ASL
evidence, four had CMRO$_2$ evidence, and three had eligible
ketone/lactate-dynamics evidence. Bipolar disorder and schizophrenia had
evidence across all five measurement classes; autism and the anxiety
disorders across four; MDD, ADHD and OCD across three; and PTSD and
clinical high risk across two. Supplementary Table~S1b lists 57
additional eligible studies located by the coverage queries and
supplementary searches; 55 of the 57 report no extractable
standardized effect, none supplies a usable pooled effect,
and one supplies the metabolic halves of two candidate pairs
(Appendix~\ref{app:evidence_pairing}). The records supporting each cell
are listed, with identifiers and, for any listed record that does not
qualify, the reason, in Supplementary Table~S6.

Four cells deserve comment. The autism FDG-PET cell is coded ``studies
exist'' although a pooled estimate exists (SMD $+0.416$, 95\% CI $-0.20$
to $1.03$): its eight pooled primaries used varying reference regions,
so the pooled value is neither in absolute units nor a comparable scalar
under the column's definition, and its standardized form does not make
it physically comparable. The three ketone/lactate cells rest on
two-state comparisons of brain lactate (a task, hyperventilation, or
lactate infusion versus rest); none measured isotopic turnover.

\subsection{Metabolism--classification pairing}
\label{app:evidence_pairing}

\begin{figure}[H]
\centering
\includegraphics[width=\linewidth]{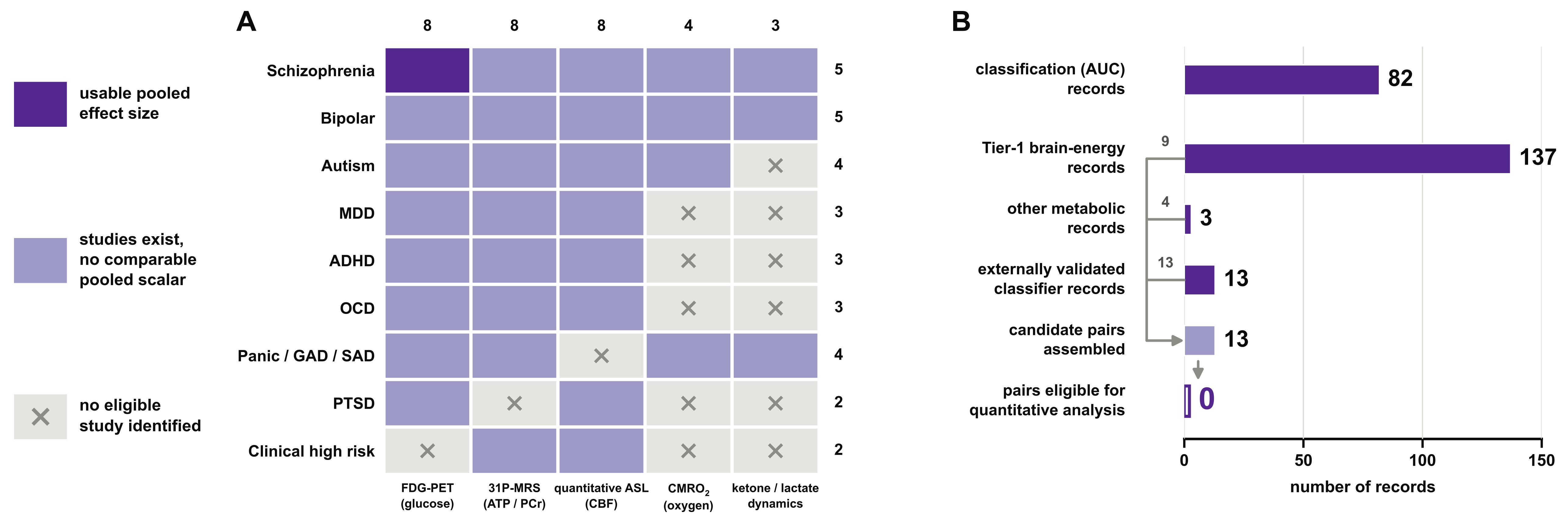}
\caption{\textbf{Coverage and comparability results of the targeted
metabolism--classification evidence map.} \textbf{A,} Coverage of five
brain-metabolism and related measurement classes across the nine
conditions of the map. Dark cells indicate that a usable pooled effect size was
available; intermediate cells indicate that eligible studies existed but
no comparable pooled scalar was available, including one cell (autism,
FDG-PET) whose pooled estimate fails the column's comparability
criteria; crossed pale cells indicate that no eligible study was
identified. Only one cell, schizophrenia FDG-PET, supplied a usable
pooled effect. The numbers above
the columns indicate the number of conditions with any evidence for the
corresponding measurement class; the numbers at right indicate the
number of measurement classes represented for each condition.
\textbf{B,} Record counts behind the attempted pairing. The map
assembled 137 Tier-1 brain-energy records (Supplementary Table~S1,
including one placeholder row; Appendix~\ref{app:evidence_pairing}) and
82 classifier-AUC records (Table~S2). The 13 candidate pairs (Table~S3)
took their classifier halves from 13 externally validated classifier
records outside Table~S2 (Table~S4). Nine of their metabolic halves are
Table~S1 records; the other four come from three further records
(``other metabolic records''): one study in Table~S1b, which supplies two
halves, and the two Alzheimer's records of
Fig.~\ref{fig:metabolic_evidence_limits} (Table~S5). Numbers
on the connectors give the halves each source supplied. None of the 13
pairs was eligible for quantitative analysis; the reason recorded for
each pair is given in Table~S3.}
\label{fig:evidence_map_coverage}
\end{figure}

The analysis distinguished the evidence pools, which establish coverage,
from the sources of the candidate pairs. The pools comprised 137
metabolic records meeting the strict Tier-1 definition, with a further
19 records in Supplementary Table~S1 (quantitative perfusion or
mixed-tracer measurements) excluded from pairing, and 82
diagnosis-versus-control AUC records from 58 studies (Supplementary
Table~S2). Besides psychiatric conditions, Table~S1 retains records for
neurological conditions and for healthy-aging and sex contrasts as
reference points; its row~1 is a placeholder noting that no quantitative
FDG-PET meta-analysis of ADHD exists, counted among the 137 but not a
study, and used neither for pairing nor as support for any coverage
cell. The classifier halves of the candidate pairs came from a
separate set of 13 externally validated, principally consortium,
classifier records from ten studies, selected to provide held-out
validation for the matched conditions (Supplementary Table~S4); none is
a Table~S2 record. Their metabolic halves came from three sources: nine
are Tier-1 records of Table~S1, two (for MDD and OCD) come from one
study in Table~S1b, and two (for Alzheimer's disease) are the two halves
of the Alzheimer's contrast in Fig.~\ref{fig:metabolic_evidence_limits}
(Supplementary Table~S5); all 13 are listed with identifiers in
Supplementary Table~S7. Matching under the rules above yielded thirteen
candidate pairs spanning nine conditions (Supplementary Table~S3;
Appendix Fig.~\ref{fig:evidence_map_coverage}B), two of them, Alzheimer's
disease and anorexia nervosa, outside the coverage grid. Candidate effect
sizes and AUCs were confirmed from the corresponding primary reports
before pairwise comparability was assessed.

None was eligible for quantitative analysis. Independence of the two
samples was not the reason: the diagnosis-level comparison this map was
designed to support joins independently estimated summaries, which is
admissible when the two halves describe the same population and the
metabolic halves share a scale. The pairs failed on commensurability, at
two levels (Supplementary Table~S3). Within pairs, the two halves
describe different populations in six cases (illness stage, chronicity
and medication, age or sex composition, or diagnostic subtype), and in
four cases the two publications are 31 years apart in diagnostic criteria
and scanner physics. Across pairs, the 13 metabolic halves fall into
six measurement classes (global absolute glucose metabolism, regional
absolute glucose metabolism, regional relative glucose metabolism,
creatine-kinase rate constants, CMRO$_2$, and mixed units), no class
holding more than three pairs. Values differ severalfold between classes
within one condition: the same Alzheimer's disease classifier estimate
pairs with metabolic effects of 2.27 and 0.84 (pairs 1 and 6), and
Fig.~\ref{fig:metabolic_evidence_limits} shows that this dependence is
general. No common scale therefore existed on which the metabolic halves
could be ordered across diagnoses. Even the two pairs with the
best-supported metabolic halves (4 and 7, schizophrenia FDG-PET pooled
over 11 and 15 studies) join metabolic pools that are chronic and
medicated (pair 4) or weighted toward such patients (pair 7) to
classifier samples of unstated chronicity. Several metabolic halves are
also imprecise, nine resting on a single primary study and three having
confidence intervals that include zero; this would have limited the
analysis but did not decide it. The 13 combinations were therefore
retained as an evidence map, and no metabolism--AUC correlation or
meta-regression was performed.

\subsection{Analysis of measurement-dependent effect sizes}
\label{app:evidence_effects}

To assess the measurement dependence of reported metabolic effect sizes,
we assembled 30 pairs of published standardized effects spanning 12
psychiatric and neurological conditions
(Fig.~\ref{fig:metabolic_evidence_limits}). Within each pair, the larger
and smaller absolute effects differed primarily in one of six respects:
spatial scope, whole-brain normalization, physical quantity, modality,
case definition, or laboratory. Nineteen pairs came from the same study
and 11 from different studies. For nonzero pairs the fold difference was
defined as
\begin{equation}
R_i=\frac{\max(|g_{i1}|,|g_{i2}|)}{\min(|g_{i1}|,|g_{i2}|)}.
\end{equation}
The median ratio was $3.9$ and the range was $1.2$--$18.2$; six pairs
contained a zero estimate and were not assigned a finite ratio. Absolute
magnitudes were plotted to emphasize scale instability, but three pairs
combined effects of opposite sign (marked in Supplementary Table~S5).
Figure~\ref{fig:metabolic_evidence_limits} should therefore be read as a
methodological stress test. It is not a pooled estimate, and it does not
imply that every condition carries a fourfold uncertainty.

Some contrasts combine measurements that are physically related but not
equivalent, and several values are derived from published summary
statistics rather than reported effect sizes. Every contrast's two
sources, identifiers, and sample sizes are given per row in
Supplementary Table~S5. The contrasts
demonstrate that spatial scope, normalization, and measurement
convention change the estimated magnitude; they are not suitable for
diagnosis ranking or formal meta-analysis.

\subsection{Interpretation and implications for study design}

This map identifies the evidence gaps that currently prevent a quantitative test of the proposed association. The literature assembled in
this targeted map did not supply matched, commensurate estimates with
which to test whether metabolic displacement predicts distributional
distinguishability or diagnostic AUC. Finding no analyzable pair says
nothing either way about the biological hypothesis itself. The most
direct test is a same-participant design: the same participants, a
common diagnosis and stage definition, calibrated energetic
measurements, connectivity and functional activity measured at
compatible scales, controlled or recorded inputs, and held-out
classification evaluated within the same sampling frame. Diagnosis-level
comparisons of independently estimated summaries, of the kind this map
was designed to support, remain admissible in principle; they require
metabolic estimates on one scale, preferably absolute, across diagnoses,
and classifier samples matched to the metabolic samples in population
and stage, neither of which the present literature supplies. Longitudinal
perturbation designs would additionally permit estimation of whether
independently measured resource availability predicts the accessible
range of distributional changes. As discussed in
Section~\ref{sec:discussion}, this tests a constraint rather than
requiring expenditure and divergence to correlate in every condition.

\end{appendices}

\end{document}